\documentclass[useAMS, usenatbib]{mn2e}
\usepackage{graphicx}
\usepackage{float}
\usepackage{placeins}
\usepackage{amssymb, amsmath}
\usepackage{epsfig}
\usepackage{epstopdf}

\newcommand\fq{$f_{\text{q}}$ }

\title[Pre-processing of infall galaxies]{The pre-processing of subhaloes in SDSS groups and clusters}

\author[Hou et al.]{Annie Hou$^{1}$, Laura C. Parker$^{1}$ $\&$ William E. Harris$^{1}$\\
$^{1}$Department of Physics $\&$ Astronomy, McMaster University, Hamilton ON L8S 4M1, Canada}

\begin{document}
 \maketitle

\begin{abstract}
We investigate pre-processing using the observed quenched fraction of group and cluster galaxies in the \citet{yang07} SDSS-DR7 group catalogue in the redshift range of $0.01 < z < 0.045$.  We categorize group galaxies as virialized, infall or backsplash and we apply a combination of the Dressler-Shectman statistic and group member velocities to identify subhaloes.  On average the fraction of galaxies that reside in subhaloes is a function of host halo mass, where more massive systems have a higher fraction of subhalo galaxies both in the overall galaxy and infall populations.  Additionally, we find that between $2 \lesssim r_{\text{200}} < 3$ the quiescent fraction is higher in the subhalo population with respect to both the field and non-subhalo populations.  At these large radii ($2 \lesssim r_{\text{200}} < 3$), the majority of galaxies ($\sim 80 \%$) belong to the infall population and therefore, we attribute the enhanced quenching to infalling subhalo galaxies, indicating that pre-processing has occurred in the subhalo population.  We conclude that pre-processing plays a significant role in the observed quiescent fraction, but only for the most massive ($M_{\text{halo}} > 10^{14.5} M_{\odot}$) systems in our sample. 
\end{abstract}
\begin{keywords}
 galaxies: groups-galaxies:dynamics-galaxies:formation
\end{keywords}

 \section{Introduction}
Observational studies of rich galaxy clusters have shown that most of the members are red early-type galaxies with little or no on-going star formation \citep{oemler74, dressler80, blanton03, balogh04, baldry06}.   While a high fraction of quiescent (i.e.\ not actively star-forming) galaxies have been observed in rich groups and clusters \citep{kauffmann04, wilman05, peng10, mcgee11, muzzin12}, recent results from observations and simulations (both numerical and semi-analytic) indicate that star formation quenching actually begins in low mass haloes with $M_{\text{halo}} \sim 10^{13} M_{\odot}$ \citep{mcgee09,bm10,george11,delucia12,wetzel12}.  Additionally, there is evidence that some cluster galaxies had their star formation quenched in groups with $M_{\text{halo}} \geq 10^{13} M_{\odot}$ prior to accretion onto the more massive cluster environment, a process often referred to as pre-processing \citep{zm98b,km08,berrier09,mcgee09,delucia12}.

While quenching has been shown to occur in low mass haloes, the significance of pre-processing is still a subject of debate.  Using $N$-body simulations \citet{berrier09} found that $70 \%$ of their cluster ($10^{14} < M_{\text{halo}} < 10^{14.6} M_{\odot}$) galaxies fell in directly from the field, while only $\sim10 \%$ fell in as members of group-sized haloes with $M_{\text{halo}} \geq 10^{13} M_{\odot}$.  Based on these results, \citet{berrier09} concluded that pre-processing did not significantly contribute to the quenched fractions observed in present-day clusters.  In contrast, both \citet{mcgee09} and \citet{delucia12} used semi-analytic models (SAMs) to show that $\sim25-45 \%$ of their simulated cluster galaxies fell in as members of systems with $M_{\text{halo}} \geq 10^{13} M_{\odot}$, where the range depends on the mass of the galaxy and the mass of the host cluster.  It should be noted that according to \citet{delucia12}, part of the discrepancy between the results of \citet{berrier09} and \citet{mcgee09} arises from differing definitions of `satellite', with the former computing fractions based on the time when a galaxy \emph{first} becomes a satellite of any halo and the latter when a galaxy becomes a satellite of the \emph{final or present-day} group or cluster. With the former definition, \citet{delucia12} find that their results are not inconsistent with those of \citet{berrier09}.  A similar analysis was carried out using $N$-body hydrodynamical simulations by \citet{bahe13}.  These authors found that $\sim 15 - 60 \%$ of galaxies in host haloes in the mass range of $10^{13.5} < M_{\text{halo}} < 10^{15.2} M_{\odot}$ had been pre-processed where the amount of pre-processing scaled with halo mass; massive haloes had a higher fraction of pre-processed galaxies \citep{bahe13}.

Thus, the results of some SAMs \citep[e.g.][]{mcgee09,delucia12} and numerical simulations \citep[e.g.][]{bahe13} predict that pre-processing can play an important role in quenching star formation, especially in massive clusters.  If the simulation predictions  of significant pre-processing  in groups and  clusters are correct  then it should be possible to  observe pre-processing  by looking at the populations  of galaxies in different environments .  The aim of this paper is to investigate the significance of pre-processing in a statistical sample of observed groups and clusters. 

Pre-processing can be investigated by studying the properties of \emph{infalling} subhalo galaxies, where a subhalo is defined as a collection of galaxies that reside in a small halo embedded within a larger parent halo.  Subhaloes can be identified by performing substructure analysis with the Dressler-Shectman (DS) Test \citep{ds88}, which can detect galaxies with kinematic properties that deviate from those of the host halo.  It should be noted that this method of identifying subhaloes differs from those used in numerical simulations.  In particular, our observational definition of subhaloes is based on identification of kinematically distinct galaxies and does not require the galaxies within the subhalo to be gravitationally bound to one another, which is usually the case for subhaloes identified in simulations.  Subhaloes, detected via the DS Test, are preferentially found on the group or cluster outskirts \citep{wb90,zm98a,hou12,dressler13} and the usual assumption is that these systems are infalling.  However, numerical simulations have shown that a large fraction of galaxies beyond the virial radius, and out to $\sim 2.5$ virial radii, have already passed through the group or cluster core \citep[i.e.\ backsplash galaxies:][]{balogh00, mamon04, gill05, mahajan11, pimbblet11,bahe13, oman13}.  Backsplash galaxies may have experienced star formation quenching due to more massive group - or cluster-related processes, and it has been suggested that much of the environmental quenching beyond the virial radius (out to $\sim2.5$ virial radii) is most likely due to the presence of a backsplash population \citep{wetzel14}.  In contrast, the infall population typically refers to galaxies that are infalling onto the host system for the first time.  Thus, any observed enhanced quenching must be a result of a transformation that occurred prior to accretion onto the host halo.

Currently, most methods of distinguishing between the virialized, infall and backsplash populations in observed groups and clusters are based on the results of simulated systems.  These classification schemes typically involve examining $|\Delta cz|/\sigma$ distributions \citep{gill05, pimbblet11} or dividing the $\Delta cz/\sigma - r_{\text{200}}$ plane into regions occupied by virialized, infall and backsplash galaxies \citep{mahajan11, oman13}.  Although each population resides in a distinct region in the full phase-space of simulated clusters, projection effects can distort these clear divisions and there is often contamination between the observed populations (to be discussed in more detail in Section \ref{popfracs}).  In addition to differences in their phase-space locations, infall and backsplash galaxies should also have subtle differences in their stellar mass distributions.  As a result of tidal disruption, backsplash galaxies should be on average less massive than infalling galaxies at the same radius \citep{gill05}, and galaxies infalling in subhaloes will typically be more massive than individual infalling galaxies \citep{mcgee09}.  Thus, in order to better probe pre-processing and environmental effects on galaxy evolution, it is important to examine the properties of virialized, infall and backsplash galaxies as a function of stellar mass and over a wide range of masses. 

Although it is well known that high-density environments, such as groups and clusters, show signs of enhanced star formation quenching with respect to the field \citep{kauffmann04, rines05, kimm09, wetzel12, woo13}, the processes that dominate this transformation are still debated.  Comparing the properties of infalling and backsplash subhalo galaxies allows us to probe the relative importance of rich group- and cluster-related processes, which are observable in the backsplash population, to pre-processing in lower mass haloes, which can be observed in the infalling subhalo population.  

In this paper we use a well-studied SDSS group catalogue to probe the properties of subhalo galaxies in groups and clusters in order to investigate the amount of pre-processing that occurs and to study the relative importance of the lower mass group environment in the evolution of galaxies.  The paper is structured as follows: in Section \ref{data_p3}, we present our group and galaxy sample and in Section \ref{identifying_sub}, we discuss how we identify subhaloes.  We compare the properties of the non-subhalo and subhalo populations, as well as compare the virialized, infall and backsplash subpopulations in Section \ref{subgals}.  Finally, in Section \ref{preprocess}, we discuss our results and present our conclusions in Section \ref{conclusions}.  Throughout this paper we assume a $\Lambda$CDM cosmology with $\Omega_{m,0} = 0.31$, $\Omega_{\Lambda,0} = 0.69$ and $H_{0} = 70$ km s$^{-1}$ Mpc$^{-1}$. 

 \section{Data}
 \label{data_p3}
Observational results have found correlations between host environments and galaxy properties; however, these correlations are more easily observed in low mass galaxies \citep[e.g.][]{peng10,carollo13,hou13}.  Thus, to fully investigate the role of pre-processing in galaxy groups, we require a large sample of group and cluster galaxies that is complete down to low stellar masses ($\log_{10}(M_{\text{star}}/M_{\odot}) \sim 9.5$), where environmental trends are expected to be more significant.  These requirements can be achieved with the \citet{yang07} Sloan Digital Sky Survey (SDSS) group catalogue.

 \subsection{The SDSS-DR7 Galaxy Catalogue}
 The galaxy magnitudes, extinctions, $k$-corrections and stellar masses are obtained from the New York University Value Added Catalogue (NYU-VAGC: \citealt{blanton05}).  The $k$-corrections and stellar masses are computed following the methodology of \citet{br07}, which assume a Chabrier Initial Mass Function (IMF).  The star formation rates (SFRs) and specific star formation rates (SSFR = SFR/$M_{\text{star}}$) are from the most recent release of the spectral reductions of \citet{brinchmann04}\footnote{http://www.mpa-garching.mpg.de/SDSS/DR7}.  The SSFRs are measured from emission lines, whenever available, or determined from the 4000 \AA \hspace{1pt} break ($D_{n}4000$) when there are no clear emission lines or in the presence of strong contamination from active galactic nuclei \citep{brinchmann04}.  It should be noted that SSFRs obtained from the $D_{n}4000$ value, which are typically values $< 10^{-12} \text{\hspace{1pt} yr}^{-1}$, are not exact measures of SSFR but should instead be taken as an upper limit.  The average $2\sigma$ errors on the SFR estimates are between 0.5 - 1.0 dex, where galaxies with higher SSFRs have lower errors \citep{brinchmann04}.

 The effects of the environment and the importance of pre-processing are probed via the quiescent fraction (hereafter \fq), where \fq is defined as 
\begin{equation}
\centering
f_{\text{q}} = \frac{\text{$\#$ galaxies with SSFR} < 10^{-11} \hspace{1pt} \text{yr}^{-1}}{\text{total $\#$ of galaxies}},
\label{fq}
\end{equation}
with  SSFR $= 10^{-11}$ yr$^{-1}$ marking the division between the main sequence of star-forming galaxies and the quiescent galaxies in the SSFR-stellar mass plane \citep{mcgee11, wetzel12}.  It should also be noted that the values in Equation \ref{fq} are weighted to account for spectroscopic incompleteness using the completeness values computed in \citet{yang07}.

An advantage of studying environmental effects via $f_{\text{q}}$, rather than with mean SSFR or SSFR distributions, is that the aforementioned uncertainty in low values of SSFR derived by \citet{brinchmann04} do not affect the results of our analysis.  To ensure that the quiescent fraction is not biased, we use a stellar mass complete sample. As a result of the magnitude limit of the SDSS survey, the stellar mass completeness limit is a function of redshift.  Therefore, in order to include low mass galaxies in our analysis we restrict the redshift range to $z \leq 0.045$, which provides us with a sample that is complete down to $3.2 \times 10^{9} M_{\odot}$.  Analysis is performed on satellite galaxies with all `central' galaxies, taken to be the most massive galaxy as identified by \citet{yang07}, removed from our sample. 

\subsection{The SDSS Group Catalogue}
\label{yang}
Our sample consists of groups and clusters identified in SDSS by \citet{yang07}.  These authors identify groups using all galaxies in the SDSS-DR7 sample brighter than the survey magnitude limit of $r \leq 17.77$ and with spectroscopic completeness $> 70 \%$.  The groups are identified with a halo-based group finder, which uses a traditional friends-of-friends algorithm to identify potential systems and then adds or removes members iteratively based on the mass of the dark matter halo and the assumption that the distribution of galaxies follows that of dark matter haloes, which is assumed to be a projected NFW profile \citep{nfw}.  The mass of the halo is determined initially from the total or characteristic luminosity ($L_{19.5}$ in \citealt{yang07}) of all the potential group members with $^{0.1}M_{r, \text{lim}} - 5\log h^{-1} \leq -19.5$, where $^{0.1}M_{r, \text{lim}}$ is the absolute magnitude limit at the redshift of the group $k$-corrected to $z = 0.1$, and a constant mass-to-light ($M/L$) ratio of $500 M_{\odot}/L_{\odot}$.  It should be noted that only for the first iteration is a constant $M/L$ ratio used; for all subsequent iterations the $M_{\text{halo}}/L_{19.5} - L_{19.5}$ relation from the previous iteration is used to determine the halo mass.  In addition, an initial velocity dispersion and size are computed from the members of the potential group.  Using this initial mass, size and dispersion, as well as an assumed NFW radial profile and a Gaussian distribution for the line-of-sight (LOS) velocities, the algorithm then adds or removes members until no further members can be added and the  $M_{\text{halo}}/L_{19.5} - L_{19.5}$ relation converges.

The \citet{yang07} group finder identifies systems that cover a wide range of masses, from isolated galaxies to rich clusters ($M_{\text{halo}} \sim 10^{15} M_{\odot}$).  These authors carried out performance tests of their halo-based group finder using a mock galaxy redshift survey made to mimic the SDSS-DR4 sample.  The performance of the group finder was characterized by the completeness ($f_{c}$), defined as the number of members identified over the total number of true group members, and the contamination ($f_{i}$), defined as the number of interloping non-members over the total number of true members.  \citet{yang07} found that the percentage of groups with $100\%$ completeness ranged from $\sim93 \%$ in low-mass groups ($10^{12.5} < M_{\text{halo}} \leq 10^{13.5} M_{\odot}$) to $60 \%$ for the most massive clusters ($10^{14.5} < M_{\text{halo}} \leq 10^{15} M_{\odot}$).  Since the majority of systems in our sample are in the low-mass halo regime, it is expected that our groups are relatively complete.  The contamination from interlopers appears to be mostly independent of halo mass.  On average $\sim 65 \%$ of the systems had no contamination at all and $\sim 85 \%$ had $f_{i} \leq 0.5$.  Interloper galaxies are typically either field galaxies or members of nearby massive groups with similar projected spatial positions but offsets along the LOS.  The impact of the number of interloping galaxies needs to be explored in detail with mock catalogues from simulations, which we reserve for future work.  However, we discuss possible implications of interloping galaxies for this work in Section \ref{identifying_sub}. 

\begin{figure}
 \centering
 \includegraphics[height = 8cm, width = 8cm]{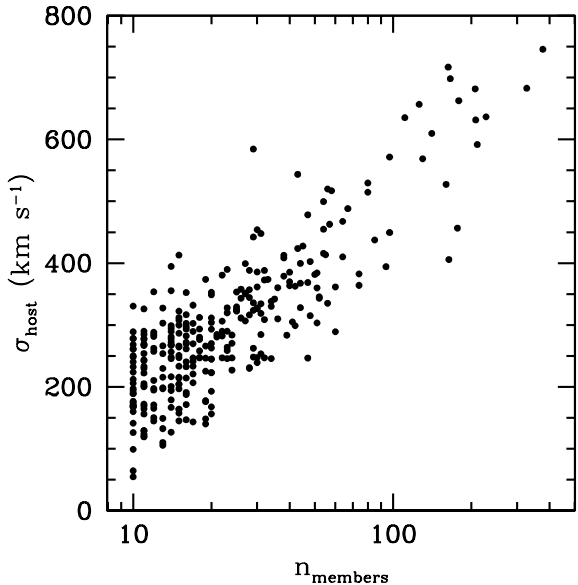}
 \caption[$\sigma_{\text{rest}}$ versus $n_{\text{members}}$ for the groups in our sample]{Rest-frame group velocity dispersion ($\sigma_{\text{rest}}$) versus group richness ($n_{\text{members}}$) for our group sample.  The value of $n_{\text{members}}$ is taken to be the number of group members \emph{after} our stellar mass cut of $3.16 \times 10^{9} M_{\odot}$ is applied.  The majority of the groups in our sample have $100 < \sigma_{\text{rest}} < 400$ km s$^{-1}$ and $10 \leq n_{\text{members}} \lesssim 50$.}
 \label{signgroups}
 \end{figure}

In this analysis, we only study systems with $n_{\text{members}} \geq 10$, which is the minimum group membership for reliable substructure analysis \citep{hou12}.  This leaves us with a total of 306 groups and 9095 member galaxies.  Additionally, while our sample contains both groups (i.e.\ systems with $10^{12} \lesssim M_{\text{halo}} \lesssim 10^{14} M_{\odot}$) and clusters (i.e.\ systems with $M_{\text{halo}} \gtrsim 10^{14} M_{\odot}$), we will refer to all systems as `groups' for simplicity.  In Figure \ref{signgroups}, we show the main properties of the groups in our sample and plot the group velocity dispersion ($\sigma_{\text{rest}}$) versus group richness ($n_{\text{members}}$) for the systems in our sample.  The dispersion ($\sigma_{\rm{rest}}$) is the observed velocity dispersion ($\sigma_{\rm{obs}}$) computed via the Gapper Estimator \citep{beers90} from all member galaxies above our stellar mass completeness limit and then corrected for redshift  (i.e.\ $\sigma_{\rm{rest}} = \sigma_{\rm{obs}}/(1 + z))$.  The group richness is taken to be the number of group members \emph{after} our stellar mass cut of $3.16 \times 10^{9} M_{\odot}$ is applied and is therefore the number of members used in the dynamical analysis presented in this work.  The majority of our sample resides in the group regime with $100 < \sigma_{\text{rest}} < 400$ km s$^{-1}$ and $10 \leq n_{\text{members}} \lesssim 50$ (Figure \ref{signgroups}).  While our sample is primarily composed of group-sized haloes, there are several rich group- and cluster-sized systems with 42 out of the 306 groups having $n_{\text{members}} > 50$.  
 
 \section{Identifying Subhaloes}
 \label{identifying_sub}
In order to investigate whether pre-processed galaxies contribute to the observed morphology-/colour-density relations \citep[e.g.][]{dressler80, blanton03, balogh04, baldry06, bamford09}, we must first identify subhaloes, which we define as a collection of galaxies that occupy the same halo within a larger host group halo.  One method of identifying subhaloes is to look for substructure, which is believed to be an indication of the recent accretion of galaxies or small groups of galaxies.  As in our previous work \citep{hou12,hou13}, we identify substructure using a modified version of the DS Test on all groups with $N_{\text{members}} \geq 10$ in our SDSS sample.  The DS Test \citep{ds88} uses both spatial and LOS velocity information to identify substructure and searches for members or groups of members with kinematic properties that deviate from those of the host group.  The DS $\delta_{i}$-deviation is computed for each galaxy as 
\begin{equation}
\centering
\delta_{i} = \left(\frac{N_{nn} + 1}{\sigma^{2}}\right)\left[\left(\overline{\nu_{local}^{i}} - \overline{\nu}\right)^{2} + \left(\sigma_{local}^{i} - \sigma\right)^{2}\right],
\label{deltai}
\end{equation}
where $1 \leq i \leq n_{\text{members}}$, $N_{nn} = \sqrt{n_{\text{members}}}$ rounded down to the nearest integer in the modified version of the test \citep{pinkney96, zm98a}, $\overline{\nu_{local}^{i}}$ and $\sigma_{local}^{i}$ are the mean velocity and velocity dispersion of the galaxy plus its $N_{nn}$ neighbours (as projected on the sky), and $\overline{\nu}$ and $\sigma$ are the mean velocity and velocity dispersion of the host group.  Galaxies with large $\delta_{i}$-values have large kinematic deviations and could indicate new group members that have yet to adopt the kinematic properties of the host group.  To determine whether a group contains significant substructure, the sum of DS deviations is computed as
\begin{equation}
\centering
\Delta = \sum_{i = 1}^{n}\delta_{i}.
\label{DS_p3}
\end{equation}
Monte Carlo methods are then used to determine the probability that the computed $\Delta$ value can be obtained from a random distribution of galaxy positions and velocities.  The probability is computed by comparing the observed $\Delta$-value to `shuffled $\Delta$-values', which are computed by randomly shuffling the observed velocities and then reassigning them to the observed member galaxy positions.  Systems with probabilities below a given confidence level (typically 1 or 5 $\%$) are identified as having significant substructure.

In our previous work, we focused on comparing groups with and without detectable substructure using the $\Delta$ statistic \citep{hou12,hou13}.  However, the goal of this work is to investigate the role of pre-processing, which requires the identification of individual subhaloes.  A simple way to identify subhaloes involves a combined analysis of the group `bubble-plot', that is a position plot of the group members where the symbols are weighted by $\exp(\delta_{i})$ and the group velocity distribution \citep{ds88, dressler13}.  In the bubble-plots the size of the symbols scale with the DS $\delta_{i}$-deviation, therefore larger symbols correspond to galaxies with larger kinematic deviations from the group average.  Subhalo candidates are identified in the bubble-plots as regions where several galaxies have similarly large symbols.  The DS $\delta_{i}$-deviation does not take into account the sign of the galaxy velocity (Equation \ref{deltai}), therefore to ensure that the galaxies are also correlated in velocity-space, \citet{dressler13} look at the velocity distribution of the subhalo candidates.  If the candidate galaxies span a small enough range in velocity ($\lesssim 1000 \text{km s}^{-1}$) then \citet{dressler13} identify these as a subhalo.

\begin{figure}
\centering
\includegraphics[width = 8cm, height = 8cm]{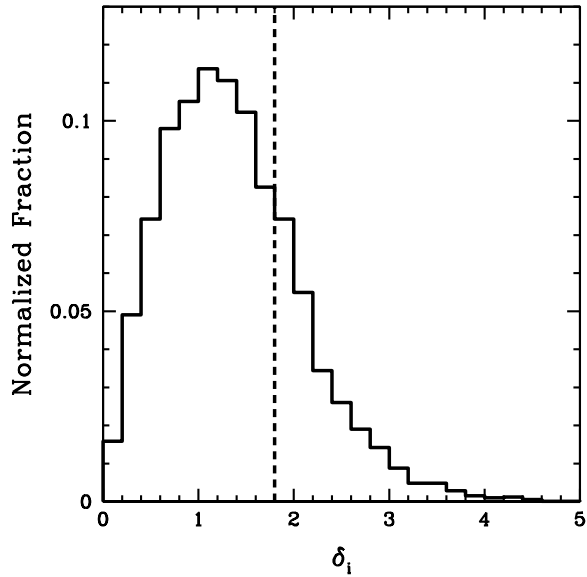}
\caption[DS $\delta_{i}$-deviation histogram]{Histogram of DS $\delta_{i}$-deviations (Eqn. \ref{deltai}) for all the SDSS satellite group members in our sample.  The dashed vertical line represents the minimum $\delta_{i}$-value required to be considered as part of a subhalo (i.e.\ $\delta_{i} \geq 1.8$.  Approximately $25 \%$ of the sample lie above $\delta_{i} = 1.8$.}
\label{deltaihist}
\end{figure}

\begin{figure*}
\centering
\includegraphics[height = 8cm, width = 8cm]{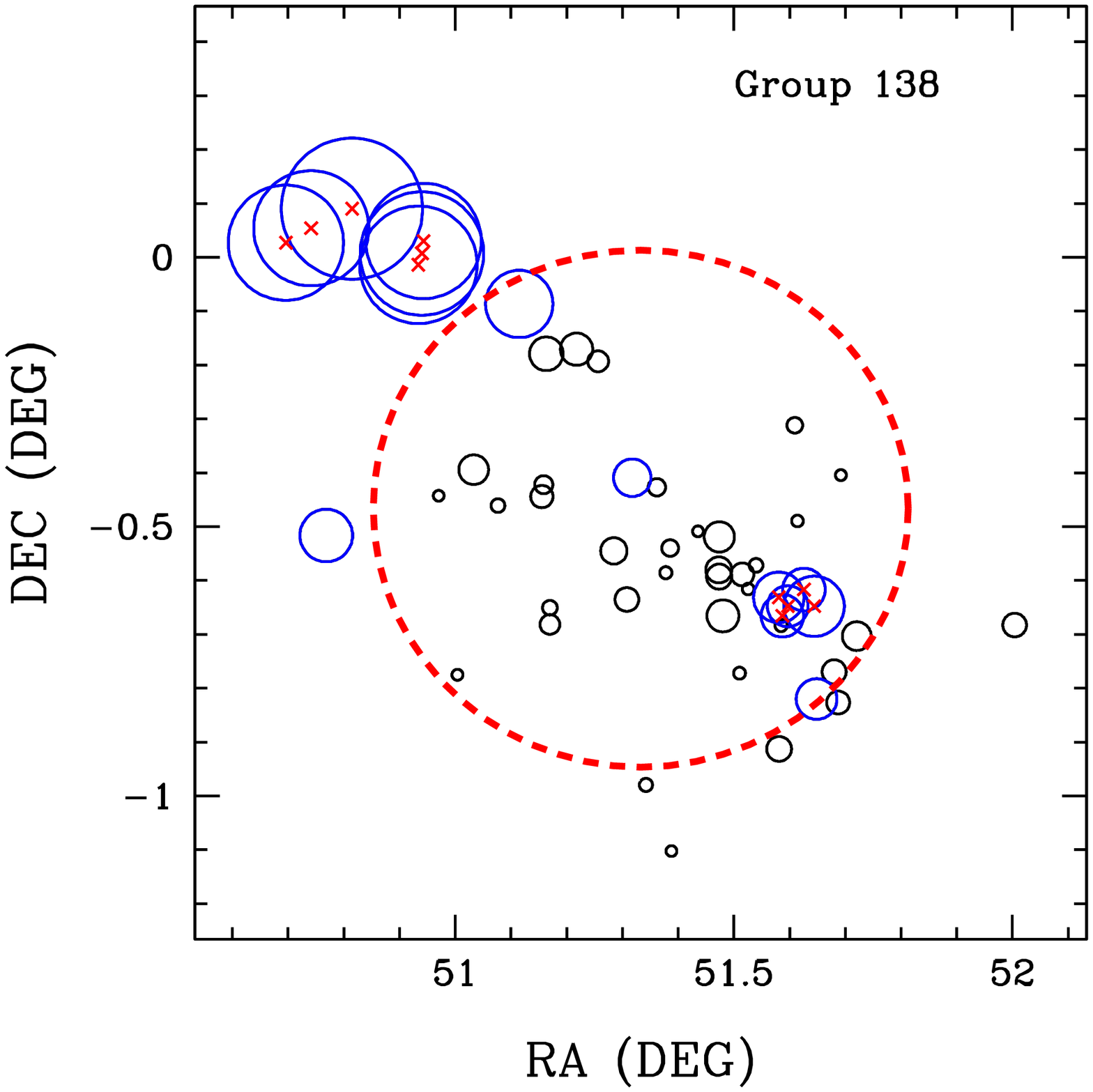}
\includegraphics[height = 8cm, width = 8cm]{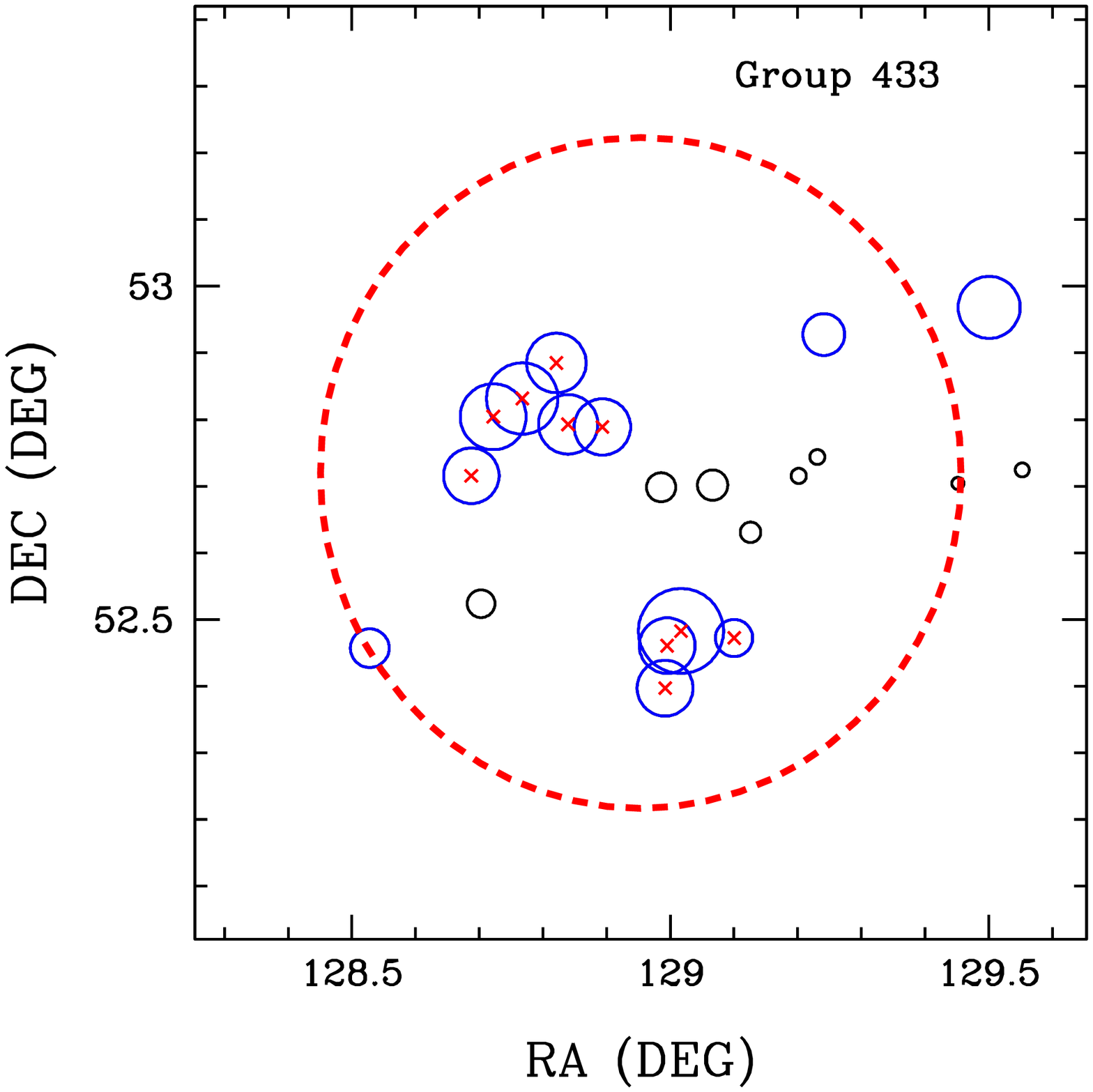}
\caption[Bubble-plot for SDSS Groups 138 and 433]{Left: Declination (DEC) versus Right Ascension (RA) for \citet{yang07} SDSS Group 138 where the symbols scale with $\exp(\delta_{i})$, often referred to as DS `bubble-plots', and larger symbols correspond to larger kinematic deviations from the host group properties.  Black symbols represent galaxies with $\delta_{i} < 1.8$, blue symbols represent galaxies with $\delta_{i} \geq 1.8$ and the red crosses indicate galaxies that have been identified as being part of subhalo by our automated subhalo finder.  Right:  Same as left except for \citet{yang07} SDSS Group 433.  The red dashed circle indicates the virial radius (i.e.\ $r_{\text{200}}$) of each system and corresponds to values of 1.27 and 1.57 Mpc for Groups 138 and 433. Subhalo galaxies identified with our automated finder (represented by red crosses) generally match those that would be identified via visual inspection (represented by blue symbols).}
\label{bubplots}
\end{figure*}

While the bubble-plots are effective in identifying subhaloes, it is not feasible to carry out visual inspection for a large sample of groups.  Therefore, we automate this process by defining subhaloes as a collection of at least three neighbouring galaxies, as projected on the plane of the sky, with $\delta_{i} \geq 1.8$ that lie within a narrow range of LOS velocities of each other. The minimum value of three neighbouring galaxies in our subhaloes corresponds to the fact that the modified version of the DS Test uses $N_{nn} = \sqrt{n_{\text{members}}}$ (rounded down to the nearest integer) to compute $\delta_{i}$ (Equation \ref{deltai}).  Since our smallest groups have $n_{\text{members}} = 10$ then at minimum $N_{nn} = 3$.  The $\delta_{i} \geq 1.8$ requirement results from the observation that the average $\delta_{i}$-value is approximately 1, which can be seen in the $\delta_{i}$-distribution shown in Figure \ref{deltaihist}.  We choose the value of $\delta_{i} = 1.8$ so that $\sim 25 \%$ of the galaxies in our sample lie above  the cut-off, but our results are not sensitive to the particular value of $\delta_{i}$ that is used.  The LOS velocity cut applied around each galaxy ensures that the candidate subhalo galaxies are not only close in projection on the sky, but also correlated in redshift space.  Since our sample includes groups that span a wide range in halo mass and group richness (Figure \ref{signgroups}), we set our LOS velocity cut equal to $\sigma_{\text{rest}}$,  which allows the velocity range for subhalo galaxies to scale with the mass of the host group.  Theoretically, subhaloes should also span a wide range in mass; however, massive subhaloes ($\sim 10^{13} M_{\odot}$) are likely only found in cluster-sized systems \citep[$\gtrsim 10^{14} M_{\odot}$:][]{mcgee09}.  A constant LOS velocity cut applied to all subhaloes would either be too restrictive for rich clusters or too relaxed for lower mass groups.  While we cannot reliably determine the masses of our identified subhaloes, due to the small number of galaxies within a given subhalo, visual inspection of our systems shows that subhaloes in lower mass groups typically only have a few member galaxies, while more massive systems can contain subhaloes with as few as three and as many as $\sim 10$ member galaxies.  Thus, our methodology reflects the expected range in subhalo masses for a given group halo mass.

We now compare our automated subhalo finder to the visual inspection methodology described in \citet{dressler13}.  In Figure \ref{bubplots}, we show bubble-plots for two example groups in our sample, \citet{yang07} SDSS Groups 138 (left) and 433 (right).  Galaxies with $\delta_{i} < 1.8$ are indicated by black symbols, galaxies with $\delta_{i} \geq 1.8$ are indicated by blue symbols and the size of the symbols scale with $\exp(\delta_{i})$.  In addition, we also indicate the galaxies identified as subhalo members using our automated algorithm (red crosses in Figure \ref{bubplots}).  In both Groups 138 and 433, our subhalo finder clearly identifies the collection of galaxies with the largest DS $\delta_{i}$-deviations (Figure \ref{bubplots}).  It appears that our automated method is able to identify the same subhaloes that a visual inspection would detect. This methodology was applied to all groups in our sample and we find that our algorithm systematically reproduces the subhalo population identified via visual inspection.

In Section \ref{yang}, we mentioned that in some cases the groups identified in \citet{yang07} suffered from contamination from interloping galaxies.  Recall that $\sim 85 \%$ of the systems in the \citet{yang07} group catalogue have anywhere between $0 - 50 \%$ contamination from interloping galaxies, which were typically galaxies on the outskirts of neighbouring groups along the LOS.  To determine the importance of contamination from neighbouring galaxies within our subhaloes, we computed the number of groups with subhaloes with close neighbours, defined as groups that are within $3r_{\text{200}}$ and $3\sigma$ (along the LOS) of another group.  We found that only $\sim 5 \%$ of our sample of groups with subhaloes had close neighbours, indicating that contamination from neighbouring groups does not have a significant effect on our identification of subhalo galaxies.

\section{Comparing subhalo and non-subhalo galaxies}
\label{subgals}
With our automated subhalo finder, we identify subhaloes in our group sample and find that in total $\sim 10 \%$ of all group galaxies reside in subhaloes.  We then construct samples of subhalo and non-subhalo galaxies and in this section we compare the galaxy properties of these two populations.

\subsection{Halo Mass Distributions}
\begin{figure*}
\centering
\includegraphics[height = 6.5cm, width = 6.5cm]{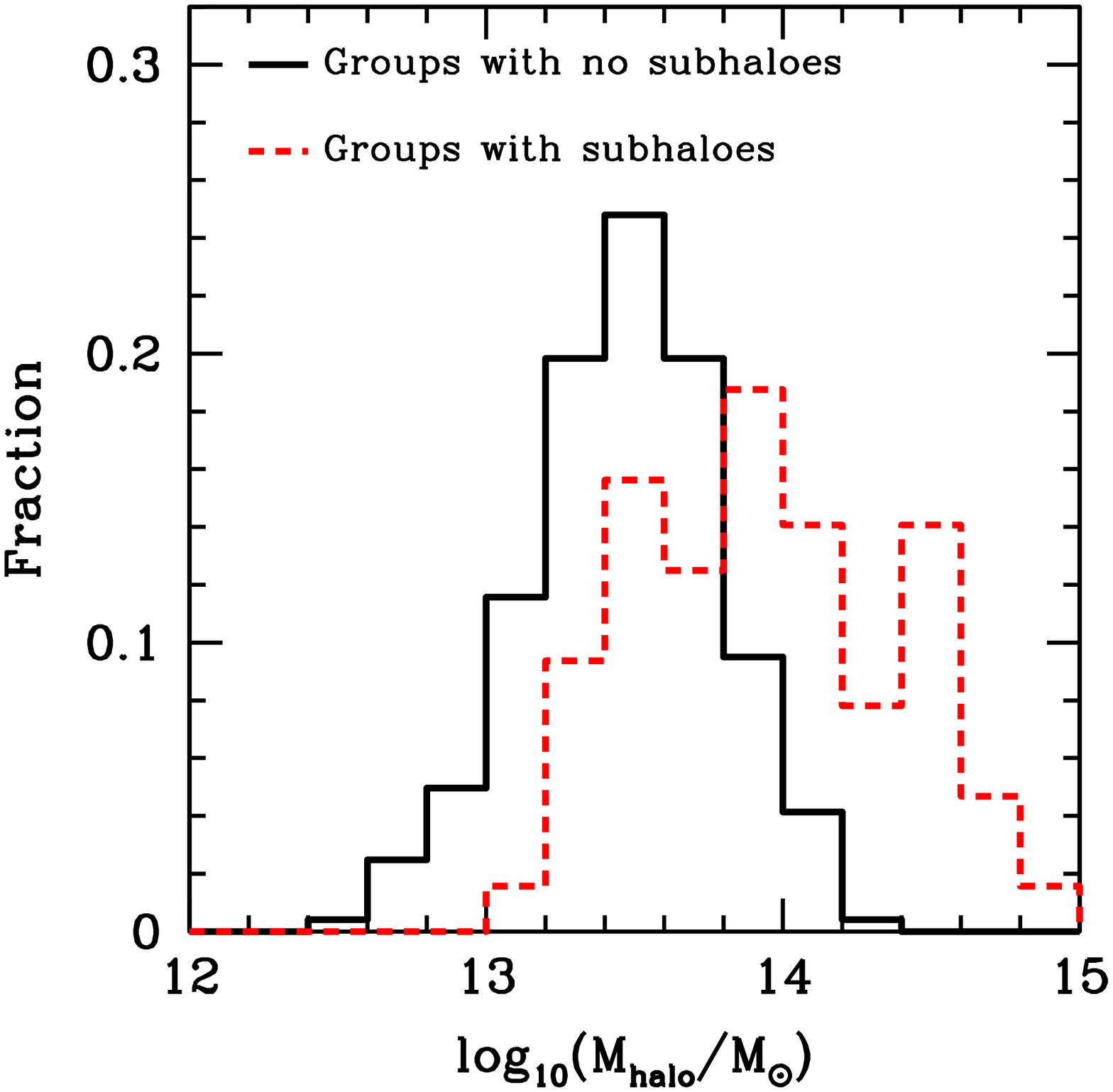}
\includegraphics[height = 6.5cm, width = 6.5cm]{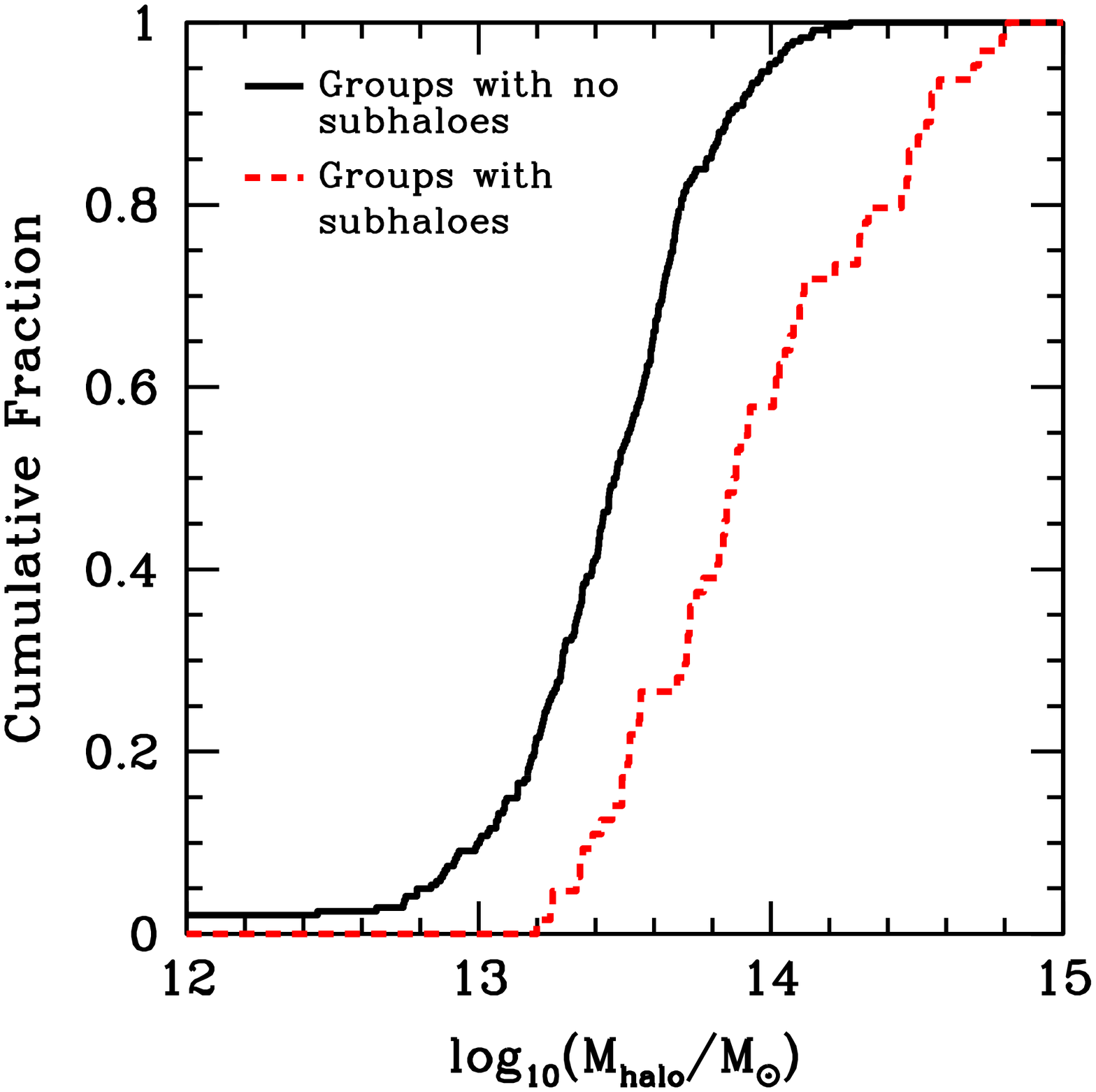}
\caption[Normalized differential and cumulative halo mass distributions for galaxies in groups with and without subhaloes]{Left: Differential halo mass distribution for galaxies in groups with no identified subhaloes (black solid line) and groups with subhaloes (red dashed line).  Right: Same as left except we plot the cumulative halo mass distributions.  It is clear that groups with subhaloes preferentially reside in more massive systems.}
\label{Mhcdf}
\end{figure*}

In Figure \ref{Mhcdf}, we show the differential (left) and cumulative (right) halo mass distributions for groups with no subhaloes (black solid line) and for groups with subhaloes identified with the methodology described in Section \ref{identifying_sub} (red dashed line).  The halo mass distributions for groups with and without subhaloes are distinct at the $> 99 \%$ confidence level based on the results of a two-sample KS Test.  It is clear that subhaloes preferentially reside in more massive systems (Figure \ref{Mhcdf}).  Almost all ($\sim 95 \%$) of the groups with no identified subhaloes have halo masses $\leq 10^{14} M_{\odot}$, while a significantly lower fraction ($\sim 60 \%$) of groups with subhaloes lie below this halo mass.  These results are generally in good agreement with results from numerical simulations and semi-analytic models (SAMs), which suggest that subhaloes are more common in more massive host groups \citep[e.g.][]{delucia12,bahe13,wetzel13b}.  We discuss the relationship between subhaloes and halo mass in more detail in Section \ref{preprocess}.

\subsection{Stellar Mass Distributions}
\label{smdists}
\begin{figure*}
\centering
\includegraphics[height = 6.5cm, width = 6.5cm]{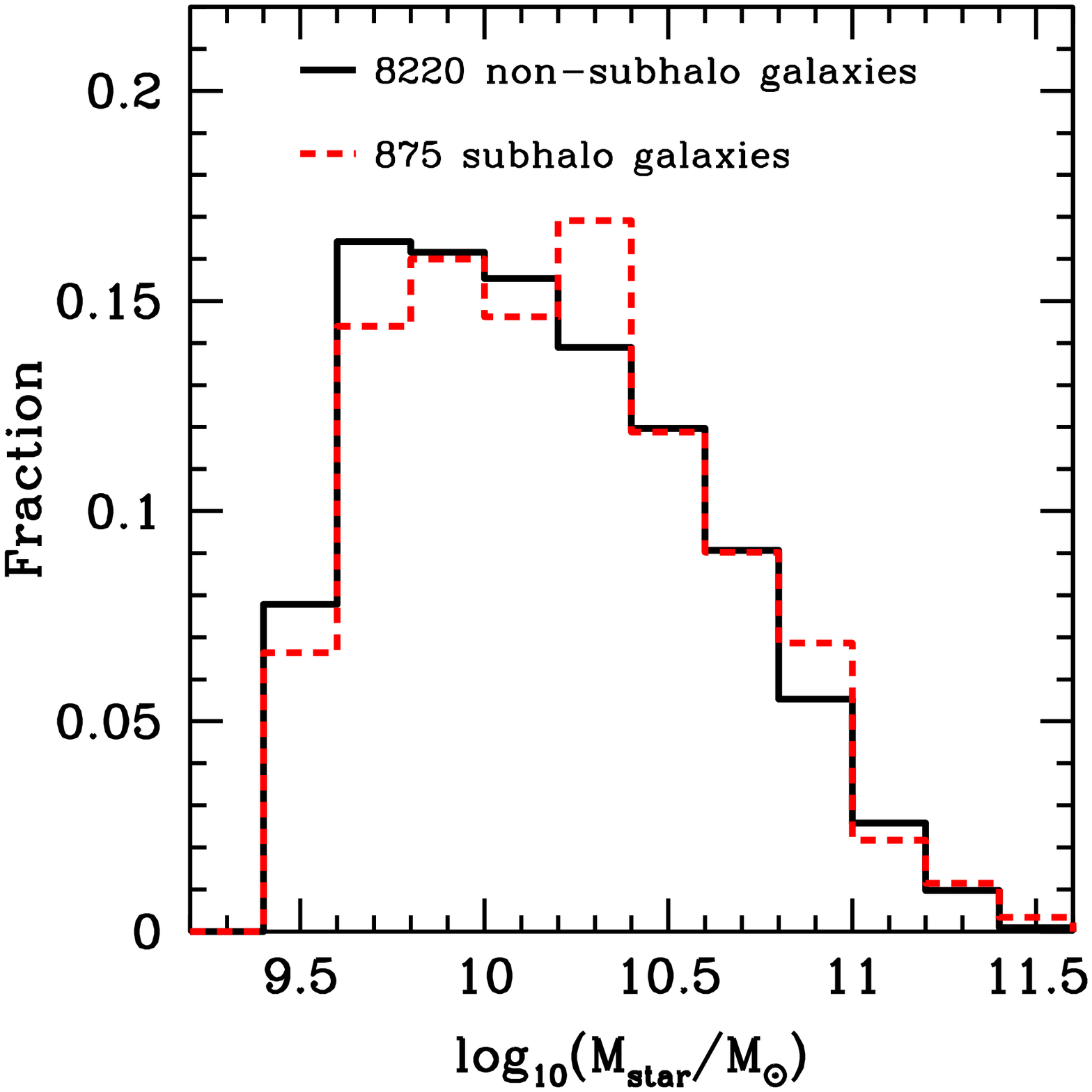}
\includegraphics[height = 6.5cm, width = 6.5cm]{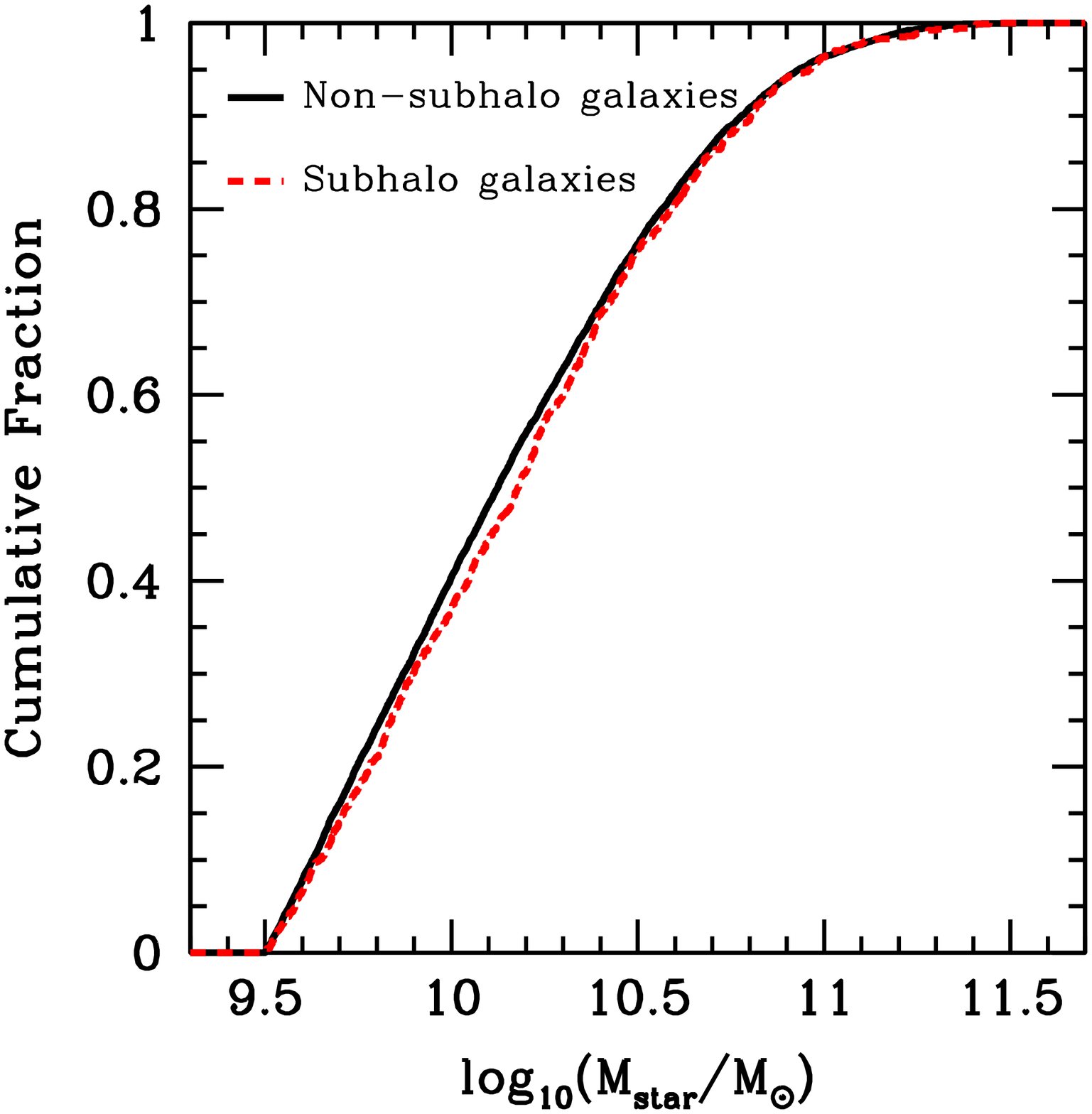}
\caption[Differential and cumulative stellar mass distributions for non-subhalo and subhalo galaxies]{Left: Normalized differential stellar mass distribution of galaxies with $M_{\text{star}} \geq 3.16 \times 10^{9} M_{\odot}$ in the non-subhalo (black solid line) and subhalo (red dashed line) populations.  Right: same as left except we plot the cumulative stellar mass distributions.  The non-subhalo and subhalo stellar mass distributions appear similar; however they are distinct a the $> 96 \%$ confidence level based on the results of a two-sample KS Test.}
\label{smhist}
\end{figure*}

We show the differential (left) and cumulative (right) stellar mass distributions for non-subhalo (black solid line) and subhalo (red dashed line) galaxies in Figure \ref{smhist}.  Although the non-subhalo and subhalo stellar mass distributions appear similar, a two-sample KS Test indicates that these two distributions likely come from distinct parent distributions at the $> 96 \%$ confidence level.  The main differences between the two populations is that the subhalo population appears to have fewer low mass ($\log_{10}(M_{\text{star}}/M_{\odot}) \lesssim 10$) galaxies and a slightly higher fraction of more massive galaxies  (Figure \ref{smhist}: left).  However, it should be noted that the differences in the stellar mass distributions of non-subhalo and subhalo galaxies are subtle and on the order of, at most, a few percent (Figure \ref{smhist}: right).  We discuss whether these differences in the stellar mass distribution affect our results in Section \ref{radtrend_subgals}.

\subsection{Radial Trends}
\label{radtrend_subgals}
In order to obtain better statistics for our analysis, we look at the stacked group properties.  Our group sample consists of groups with varying sizes and so we show radial trends as a function of $r_{\text{proj.}}/r_{\text{200}}$, where $r_{\text{proj.}}$ is the projected group-centric radius and $r_{\text{200}}$ is defined as the radius within which the average density is 200 times the critical density of the Universe.  We use an approximation of the phsyical virial radius as defined in \citet{carlberg97}
\begin{equation}
\centering
r_{200} = \frac{\sqrt{3}\sigma_{\rm{rest}}}{10H(z)},
\label{r200_p3}
\end{equation}
where $H(z) = H_{0}\sqrt{\Omega_{m,0}(1 + z)^{3} + \Omega_{\Lambda,0}}$. 

\subsubsection{Radial Distributions}
\begin{figure*}
\centering
\includegraphics[height = 6.5cm, width = 6.5cm]{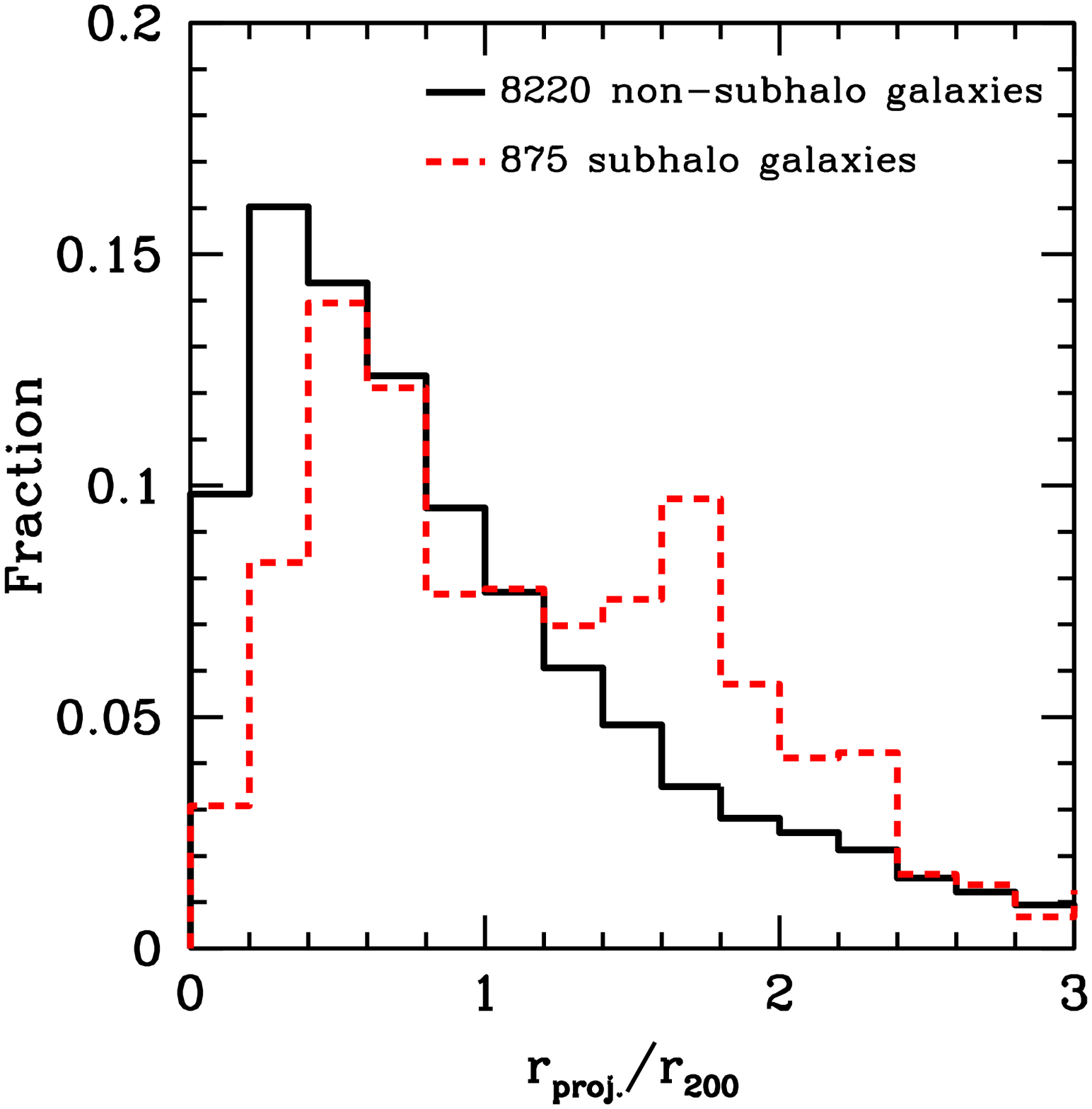}
\includegraphics[height = 6.5cm, width = 6.5cm]{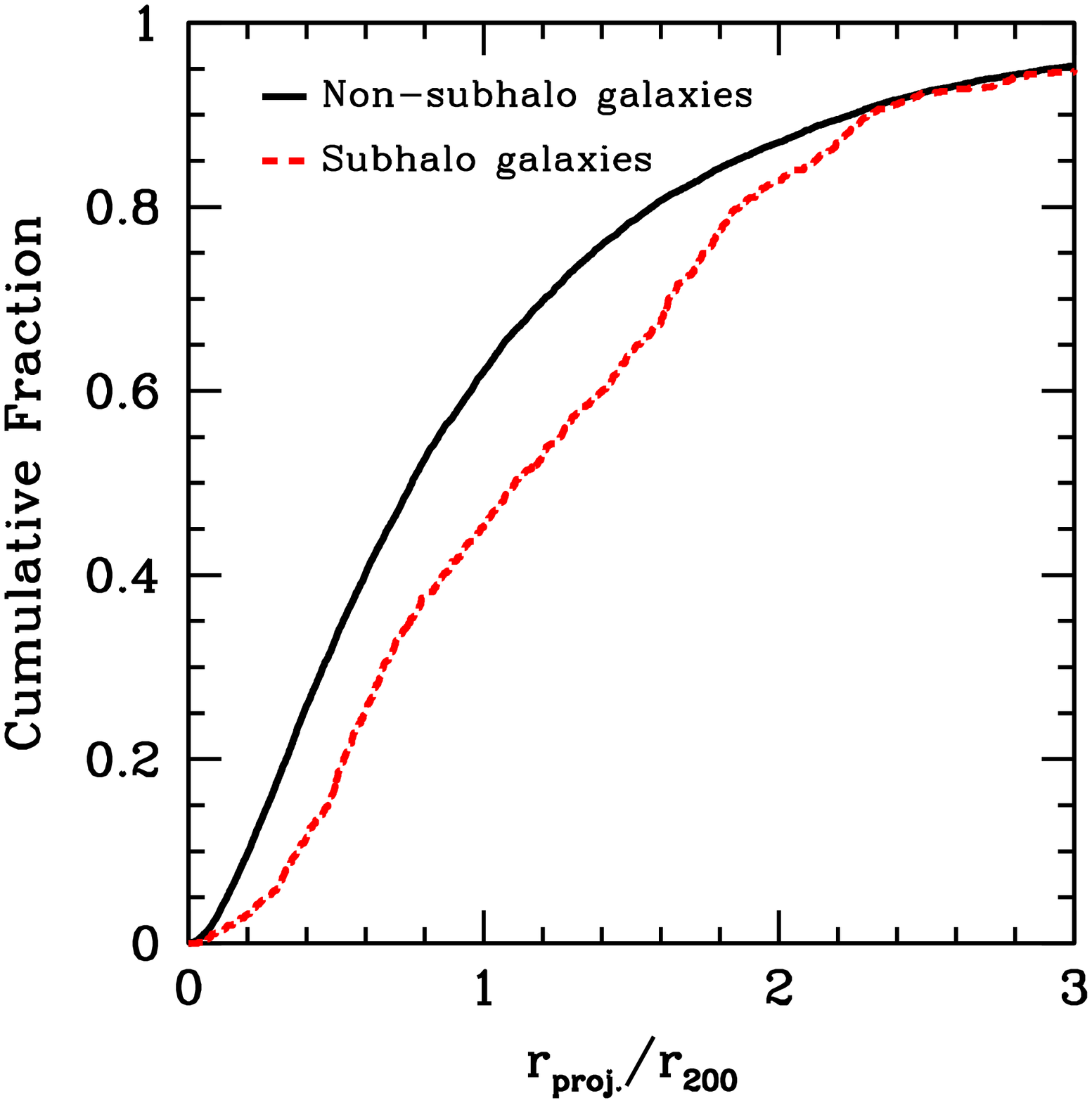}
\caption[Differential and cumulative radial distributions for non-subhalo and subhalo galaxies]{Left: Differential radial distributions of galaxies with $M_{\text{star}} \geq 3.16 \times 10^{9} M_{\odot}$ in the non-subhalo (black solid line) and subhalo (red dashed line) populations. Right: same as left except we plot the cumulative radial distributions.  Subhalo galaxies are preferentially located on the group outskirts.}
\label{radhist_p3}
\end{figure*}

In Figure \ref{radhist_p3}, we show the differential (left) and cumulative (right) group-centric radial distributions for galaxies in the non-subhalo (black solid line) and subhalo (red dashed line) populations in our stellar mass complete sample.  From Figure \ref{radhist_p3}, we see that the two distributions differ and results from a two-sample KS Test confirm that the subhalo and non-subhalo radial distributions come from different parent distributions at the $> 99 \%$ confidence level.  Additionally, we find that subhalo galaxies are preferentially found at larger radii when compared to non-subhalo galaxies (Figure \ref{radhist_p3}).  The majority ($\sim 60 \%$) of the galaxies in the non-subhalo population reside within the virial radius and the fraction of galaxies decreases with increasing radius.  In contrast, there appears to be a dearth of subhalo galaxies close to the group core with $\sim 60 \%$ of subhalo galaxies found beyond the virial radius. 

\subsubsection{Quiescent fraction versus radius}
\label{fqradsec}
\begin{figure*} 
\centering
\includegraphics[width = 14cm, height = 14cm]{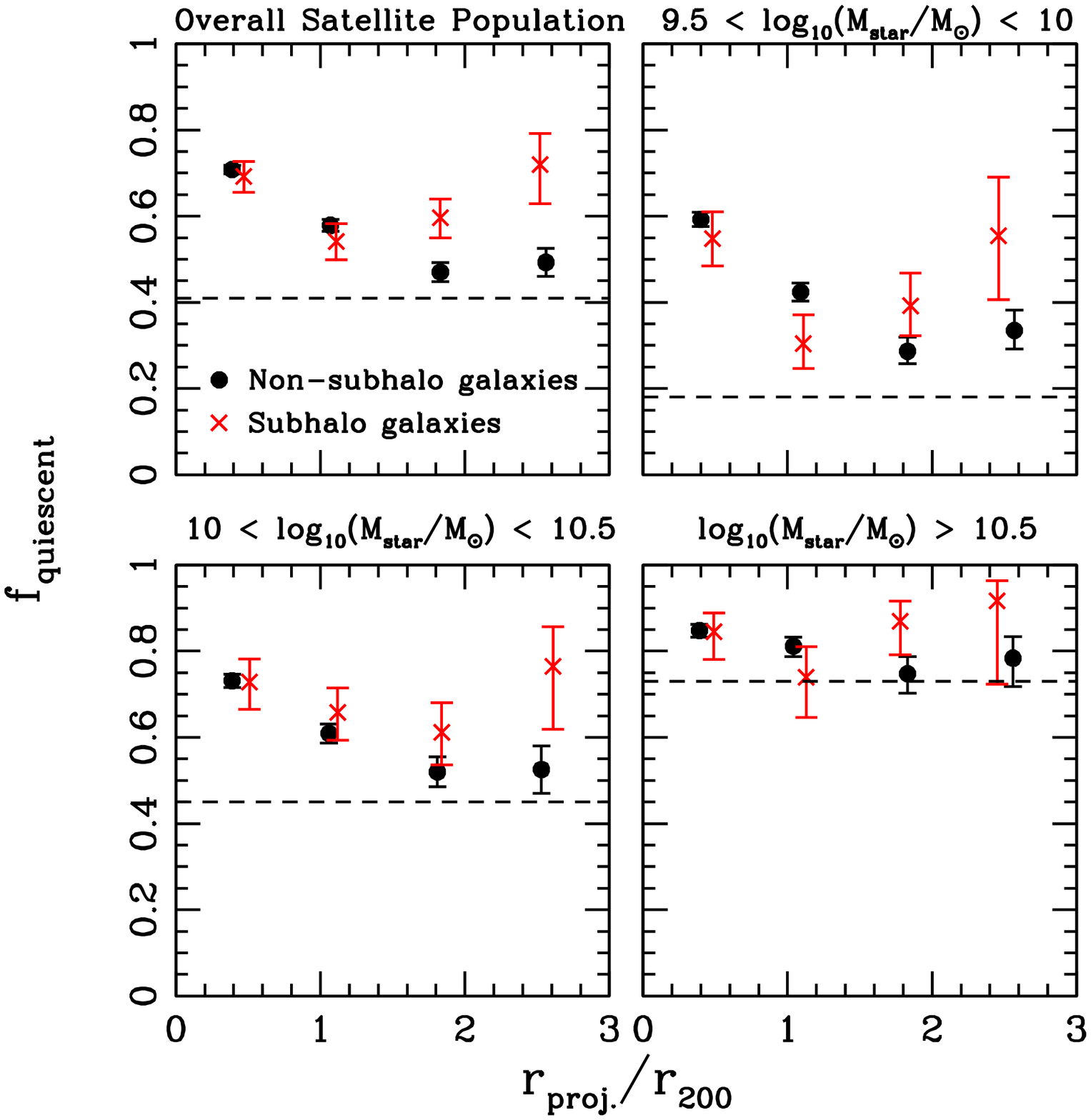}
\caption[Quiescent fraction versus projected radius for non-subhalo and subhalo galaxies in narrow bins of stellar mass]{Top-left: Quiescent fraction ($f_{\text{q}}$) versus $r_{\text{proj.}}/r_{\text{200}}$ for all satellite non-subhalo galaxies (black circles) and all satellite subhalo galaxies (red crosses).  Top-right: same as top-left except only for the low mass ($9.5 < \log_{10}(M_{\text{star}}/M_{\odot}) < 10$) galaxies.  Bottom-left: same as top-left except for intermediate mass ($10 < \log_{10}(M_{\text{star}}/M_{\odot}) < 10.5$) galaxies.  Bottom-right: same as top-left except for high mass ($\log_{10}(M_{\text{star}}/M_{\odot}) > 10.5$) galaxies.  The data are plotted at the mean value of each radial bin, which has a width of 0.75 $r_{\text{200}}$.  The dashed horizontal black line represents the quiescent fraction in isolated field galaxies in the aforementioned stellar mass bins.  For the overall satellite population (top-left panel) the field quiescent fraction is weighted to match the group galaxy stellar mass distribution.  Errors are computed following \citet{cameron11}.  In general, the group galaxies (both non-subhalo and subhalo) have higher quiescent fractions than observed in the field.  Additionally, on the group outskirts, \fq is higher in subhalo galaxies with respect to non-subhalo galaxies, indicating that enhanced quenching has occurred.}
\label{fqrad_subgals}
\end{figure*}

Differences between subhalo and non-subhalo galaxies can be probed by looking at their SSFRs via the quiescent fraction ($f_{\text{q}}$).  In Figure \ref{fqrad_subgals}, we show \fq versus $r_{\text{proj.}}/r_{\text{200}}$ for non-subhalo (black circles) and subhalo (red crosses) galaxies in our entire sample of satellite galaxies (top-left panel), for low mass satellites ($9.5 < \log_{10}(M_{\text{star}}/M_{\odot}) < 10$: top-right panel), for intermediate mass satellites ($10 < \log_{10}(M_{\text{star}}/M_{\odot}) < 10.5$: bottom-left panel) and high mass satellites ($\log_{10}(M_{\text{star}}/M_{\odot}) > 10.5$: bottom-right panel).  The group-centric radius covers a range between $0 < r_{\text{proj.}}/r_{\text{200}} < 3$ and the data are plotted at the mean value of each bin, which have widths of 0.75 $r_{\text{200}}$.  The dashed horizontal black line corresponds to the observed quiescent fraction in the field, where field galaxies are taken to be the isolated galaxies in the \citet{yang07} catalogue.

In the top-left panel of Figure \ref{fqrad_subgals} we see that for the overall group galaxy population \fq is significantly higher than in the field at all radii, indicating that group galaxies experience environmental star formation quenching out to at least three virial radii. It should be noted that the quiescent fraction for the overall satellite population is weighted to match the group galaxy stellar mass distribution for a better comparison between the field and group samples.  This result is in agreement with previous observations that also find a higher quiescent fraction in groups with respect to the field as far out as $\sim 5r_{\text{200}}$ \citep[e.g.][]{ vdl10,bahe13,wetzel14}.  Comparing group and field galaxies at a given stellar mass, we find enhanced quenching in low and intermediate mass group galaxies at almost all radii.  For high mass galaxies, the non-subhalo population shows enhanced quenching at small radii ($r < 1.5 r_{\text{200}}$), but have \fq values similar to field on the group outskirts (Figure \ref{fqrad_subgals}: bottom-right panel).  The high mass subhalo galaxies show enhanced quenching, with respect to the field, closer to the group core ($r < 0.75 r_{\text{200}}$) and just beyond the virial radius ($1.5 < r < 2.25 r_{\text{200}}$).

As a function of radius, we see that for non-subhalo galaxies at all stellar masses the general trend is that \fq decreases with increasing group-centric radius within $\sim1.5r_{\text{200}}$ and then flattens on the group outskirts (Figure \ref{fqrad_subgals}).  The subhalo galaxy population shows a different radial trend from non-subhalo galaxies, in both the overall satellite population and at fixed stellar mass, where \fq decreases with increasing radius within $\sim1.5r_{\text{200}}$ but then appears to increase at large radii (Figure \ref{fqrad_subgals}).  These results indicate that on the group outskirts subhalo galaxies have experienced enhanced star formation quenching, with respect to the non-subhalo population.  Within the virial radius the quiescent fractions in non-subhalo and subhalo galaxies are similar; however, at large radii \fq is higher in the subhalo population at all stellar masses, although a statistically significant difference is only observed in low and intermediate mass satellites beyond two virial radii (Figure \ref{fqrad_subgals}).  

In Figure \ref{smhist} we found that the non-subhalo and subhalo stellar mass distributions differed slightly, where more massive galaxies are preferentially found in subhaloes.  The quiescent fraction has also been shown to correlate with stellar mass where more massive galaxies typically have higher values of \fq \citep[e.g.][]{kimm09,wetzel12,hou13,woo13}.  While this could potentially affect the overall galaxy population (Figure \ref{fqrad_subgals}: top-left panel), the difference in the stellar mass distributions of non-subhalo and subhalo populations is very small.  More importantly, we still observe higher \fq at fixed stellar in the subhalo population (Figure \ref{fqrad_subgals}), which suggests that environmental effects contribute to the enhanced quenching.

\subsection{Separating virialized, infalling and backsplash galaxies}
\label{popfracs}

In Section \ref{radtrend_subgals}, we showed that the $f_{\text{q}} - r_{\text{200}}$ trend for subhalo and non-subhalo galaxies differed on the group outskirts.  As a result, one might naively assume that the identified subhaloes are infalling low mass groups.  However, numerous simulations have shown that backsplash galaxies can extend as far out as 2-3 virial radii \citep{balogh00, mamon04, gill05, oman13}.  Additionally, results from semi-analytic models have shown that subhaloes can survive, that is maintain the kinematic properties of the subhalo, for several orbits within the host group potential \citep{tb04}.  Therefore, our goal is to distinguish between the infall and backsplash populations in order to see if our subhaloes are composed of pre-processed infalling galaxies or backsplash galaxies that have been quenched after passing through the host group core.  Although it is notoriously difficult to disentangle the infall and backsplash populations, there has been significant effort in recent years using simulations and mock catalogues to develop a classification scheme from observable properties \citep{gill05, mahajan11, pimbblet11}.  We make use of a combination of these schemes to identify virialized, infall and backsplash galaxies in our sample.

One way to determine if our groups contain infall and backsplash satellites is to look at the distribution of the galaxy velocities ($\Delta cz$) as a function of the group velocity dispersion ($\sigma$).  Backsplash galaxies will have been slowed due to dynamical friction within the group core and will therefore have low $|\Delta cz|/\sigma$ values at fixed radius \citep{gill05}.  In contrast, infalling galaxies can have a wide range of velocities depending on their orbital parameters; though galaxies with high velocities (i.e. $|\Delta cz|/\sigma \gtrsim 1$) are likely all infalling.  Using $N$-body simulations, \citet{gill05} showed that backsplash galaxies have a narrow centrally peaked $|\Delta cz|/\sigma$ distribution, while infalling satellites have a broader distribution with a non-zero peak.  While it is difficult to separate infall and backsplash galaxies from an observed, and therefore projected, $|\Delta cz|/\sigma$ distribution, \citet{pimbblet11} found that by binning the $|\Delta cz|/\sigma$ histogram into narrow bins of radius, it is possible to identify regions where infalling galaxies dominate.  More specifically, these authors found that bimodality and/or a shift in the peak of the $|\Delta cz|/\sigma$ distribution to larger values indicated a large infall population.  

\begin{figure*}
\centering
\includegraphics[scale = 0.4]{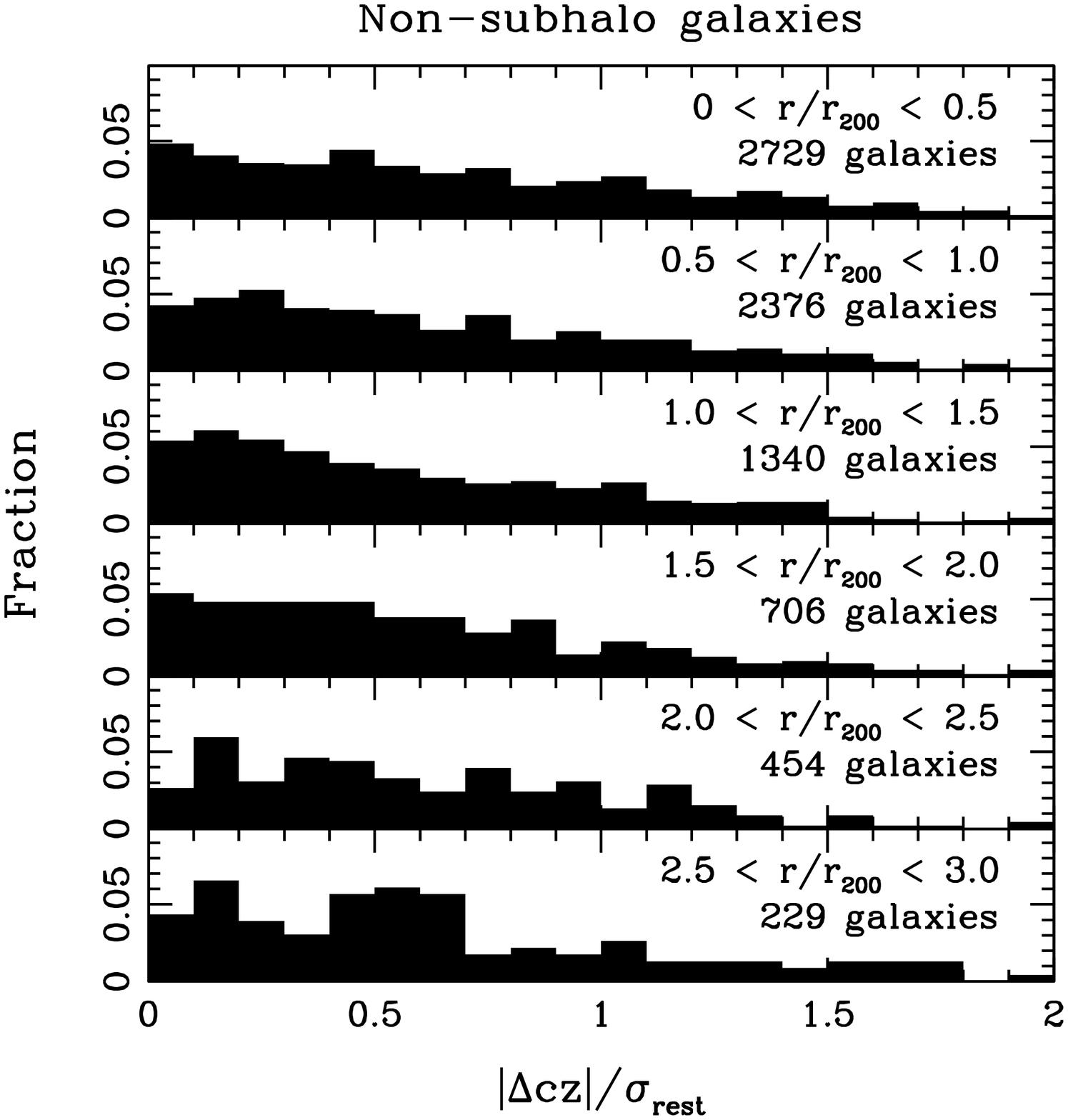}
\includegraphics[scale = 0.4]{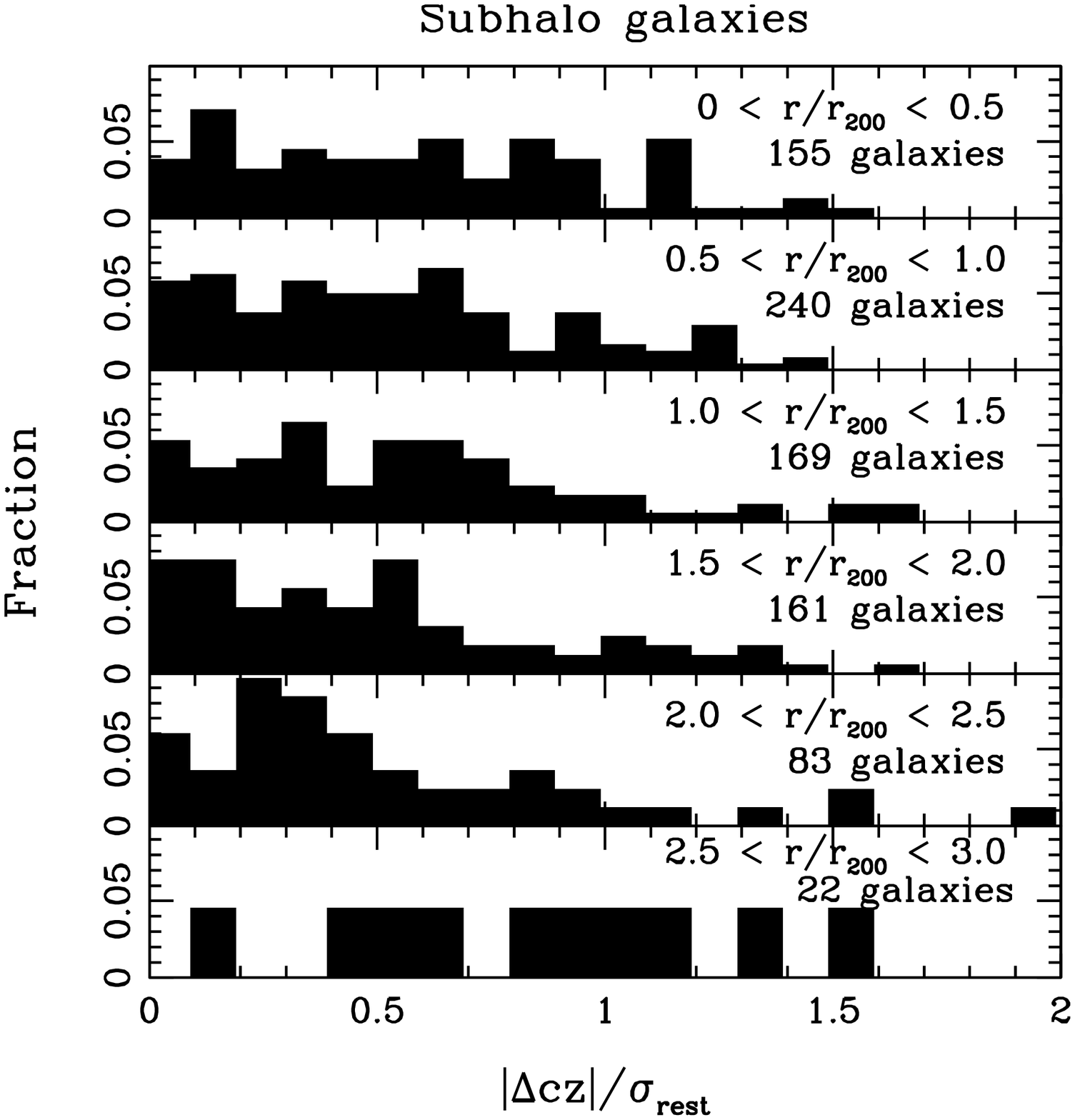}
\caption[$|\Delta cz|/\sigma_{\text{rest}}$ histograms for non-subhalo and subhalo galaxies]{$|\Delta cz|/\sigma_{\text{rest}}$ histograms for non-subhalo (left) and subhalo (right) galaxies in the $0 < r_{\text{200}} < 3$ range in bins of $0.5 r_{\text{proj}}/r_{\text{200}}$.  For each radial bin we list the number of galaxies in each population.}
\label{velhist_sub}
\end{figure*}

In Figure \ref{velhist_sub}, we show the $|\Delta cz|/\sigma_{\text{rest}}$ histograms in narrow radial bins for non-subhalo (left) and subhalo (right) galaxies and also list the number of galaxies in each bin.  For the non-subhalo galaxies we see that for almost all radial bins the $|\Delta cz|/\sigma_{\text{rest}}$ distribution is broad and generally centrally peaked, indicating a mixed population of virialized (for galaxies with $r < r_{\text{200}}$), infall and backsplash galaxies out to  $2 r_{\text{200}}$ \citep{gill05}.  Only on the outskirts ($2\leq r_{\text{200}} \leq 3.0$) are there signs of a large infall population, indicated by the emergence of a second peak at  $|\Delta cz|/\sigma_{\text{rest}} \sim 0.5$.  In contrast, the subhalo galaxies (Figure \ref{velhist_sub}: right) show signs of a strong infall population just beyond the virial radius and out to $3 r_{\text{200}}$.  The $|\Delta cz|/\sigma_{\text{rest}}$ distributions for galaxies between $1 - 2.5 r_{\text{200}}$ either show bimodality or an offset peak, which are both indications of a dominant infall population \citep{gill05, pimbblet11}.  There are too few subhalo galaxies in the $2.5 < r < 3 r_{\text{200}}$ bin to comment on the shape of the $|\Delta cz|/\sigma_{\text{rest}}$ distribution; however, it is clear that most of the galaxies have relatively high velocities and are likely infalling.

The $|\Delta cz|/\sigma$ distributions for the subhalo galaxies shown in Figure \ref{velhist_sub} indicate the presence of a dominant infall population; however, it is not possible to distinguish between infall and backsplash galaxies from these histograms alone.  Using $N$-body simulations, both \citet{gill05} and \citet{oman13} showed that in 6-d phase-space ($x$, $y$, $z$, $v_{x}$, $v_{y}$ and $v_{z}$), the regions occupied by each population are for the most part distinct.  However, once this phase-space is collapsed into observables (i.e.\ $x$, $y$ and $v_{z}$), projection effects tend to fill out much of the empty phase-space that separated the populations.  While there is no ideal method to distinguish between infall and backsplash galaxies in observed groups, there are ways to roughly approximate regions occupied by either population.  \citet{mahajan11} found that the fraction of virialized, infalling and backsplash galaxies occupied distinct regions in the $v_{r}/V_{v} - r/R_{v}$ plane, where $r$ and $v_{r}$ are the radial phase-space coordinates, $V_{v}$ is the group or cluster velocity dispersion and $R_{v}$ is the virial radius of the system.  To distinguish between the different galaxy populations, \citet{mahajan11} make the following cut  

\begin{equation}
\centering
\frac{v_{r}}{V_{v}} =  - 1.8 + 1.06\left(\frac{r}{R_{r}}\right),
\label{infallcut}
\end{equation}
to separate backsplash and infall galaxies and a cut at one virial radius to separate virialized and infall galaxies.  Within the viral radius there is an additional cut to separate virialized and infall galaxies, which is the mirror slope of Equation \ref{infallcut}.  It should be noted that while these cuts are based on the full 6-d phase-space data, the density contours for the virialized, infall and backsplash populations occupy similar regions in projected space, though with significant overlap, and therefore contamination between the populations \citep{mahajan11}.   While the distinct regions are not as clear in projected space, the divisions made by Equation \ref{infallcut} allow us to approximately distinguish between infall and backsplash subhaloes, rather than assuming all subhaloes are infalling.  Therefore, we apply a cut analogous to Equation \ref{infallcut}, except $v_{r}/V_{v}$ is replaced by the observable quantity $\Delta cz/\sigma_{\text{rest}}$ and $r/R_{r}$ is replaced by $r_{\text{proj.}}/r_{\text{200}}$.  Additionally, the aforementioned classification scheme is one of five models tested by \citet{mahajan11}.  While we elect to use the best-fitting scheme, as determined by \citet{mahajan11}, it should be noted that the fraction of backsplash galaxies can change by as much as $\sim 20 \%$ depending on the classification scheme used.

\begin{figure}
\centering
\includegraphics[width = 8cm, height = 8cm]{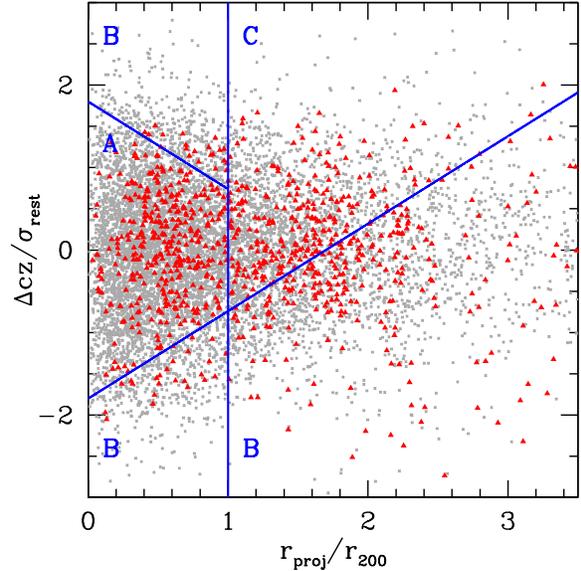}
\caption[$\Delta cz/\sigma_{\text{rest}}$ versus projected radius for non-subhalo and subhalo galaxies]{$\Delta cz/\sigma_{\text{rest}}$ versus $r_{\text{proj.}}/r_{\text{200}}$ for non-subhalo (grey crosses) and for subhalo (red triangles) galaxies.  The blue slopes indicate the line dividing galaxy populations (Equation \ref{infallcut}) as defined in \citet{mahajan11}.  Region A denotes virialized galaxies, regions B denote infalling galaxies and region C denotes backsplash galaxies.}
\label{velrad}
\end{figure}

In Figure \ref{velrad}, we plot $\Delta cz/\sigma_{\text{rest}}$ versus $r_{\text{proj.}}/r_{\text{200}}$ for our population of non-subhalo galaxies (gray crosses) and subhalo galaxies (red triangles).  As in \citet{mahajan11}, we divide the $\Delta cz/\sigma_{\text{rest}} - r_{\text{200}}$ plane into regions of virialized (Region A), infalling (Regions B) and backsplash (Region C) with Equation \ref{infallcut} and a cut at $r_{\text{proj.}}/r_{\text{200}} = 1.0$.  Both subhalo and non-subhalo galaxies occupy all three regions of Figure \ref{velrad}, though there are some visible differences between the two populations.  In particular, there are few subhalo galaxies close to the group core (also seen in Figure \ref{radhist_p3}) and there appears to be an excess of subhalo galaxies, with respect to the non-subhalo population, in the bottom-right hand corner of Figure \ref{velrad}.  This area corresponds to the region occupied by only infalling galaxies in the full 6-D phase space diagram shown in \citet{mahajan11}.  

\begin{figure}
\centering
\includegraphics[width = 8cm, height = 8cm]{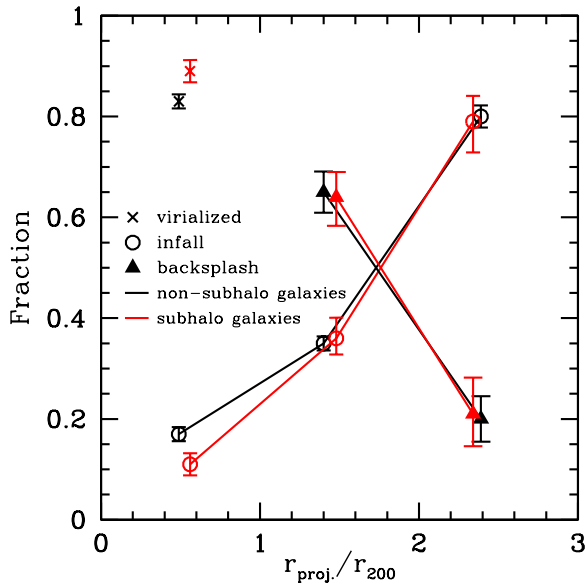}
\caption[Virialized, infall and backsplash fractions for non-subhalo and subhalo galaxies]{Fraction of virialized (crosses), infalling (circles) and backsplash (triangles) for the non-subhalo (black symbols and lines) and subhalo (red symbols and lines) populations as a function of projected group-centric radius.  The data are plotted at the mean value of each radial bin, which has a width of 1 $r_{\text{200}}$.  Errors include both $\sqrt{N}$ counting statistics and the uncertainty in the classification scheme given in \citet{mahajan11}, which are typically $1 \%$ for the virialized and infall populations and $4 \%$ for the backsplash population.  It should be noted that systematic errors, based on the method of classification, are not included.}
\label{fracpops}
\end{figure}

With the divisions shown in Figure \ref{velrad}, we separate our sample into virialized, infalling and backsplash galaxies.  In Figure \ref{fracpops} we show the fraction of virialized (crosses), infalling (circles) and backsplash (triangles) for the non-subhalo (black symbols and lines) and subhalo (red symbols and lines) populations as a function of projected group-centric radius.  The data are plotted at the mean value of each radial bin, which has a width of $r_{\text{200}}$.  The errors quoted in Figure \ref{fracpops} include both $\sqrt{N}$ counting statistics errors and the classification scheme errors stated in \citet{mahajan11}, which are typically $1 \%$ for the virialized and infall populations and $4 \%$ for the backsplash population.  However, the statistical errors quoted by \citet{mahajan11} underestimate the true uncertainty in computing the fraction of virialized, infall and backsplash galaxies, as these authors show that the fractions can change by a significant amount ($\sim 20 \%$) depending on how the galaxies are classified. 

From Figure \ref{fracpops}, we see that at fixed radius the non-subhalo and subhalo populations have very similar subpopulations.  In particular, within the virial radius both samples are dominated by virialized galaxies; though the non-subhalo population has a slightly higher fraction of virialized galaxies.  Focusing now on the infall and backsplash populations beyond the virial radius, we see that between $1 < r_{\text{200}} < 2$ roughly two-thirds of the galaxies are part of the backsplash population.  However, in the $2 < r_{\text{200}} < 3.0$ regime, the majority of the satellites reside in the infall population ($\sim 80 \%$).  Even if we include the $\sim 20 \%$ range in systematic uncertainties in classifying the infall and backsplash population fractions, we find that infall galaxies still dominate the galaxy population at large radii ($\gtrsim 2r_{\text{200}})$.

In general, our observed fractions of infall and backsplash galaxies are in relatively good agreement with values predicted from some $N$-body simulations \citep{gill05,bahe13}.  However, we do observe a higher fraction of backsplash galaxies, at all radii, than predicted by the numerical simulations of \citet{wetzel14}.  Although, if we take into account systematic uncertainties based on the method of classification, our observed fractions are in closer agreement with those of \citet{wetzel14}.   

The results of Figure \ref{fracpops} indicate that the enhanced quenching in subhalo galaxies seen beyond the virial radius in Figure \ref{fqrad_subgals} results from a combination of pre-processed infalling galaxies and backsplash galaxies that may have had their star formation quenched via processes related to the host group.  However, the largest, and most statistically significant, difference in quiescent fraction between non-subhalo and subhalo galaxies occurs at large radii ($\gtrsim 2r_{\text{200}}$: Figure \ref{fqrad_subgals}).  Between 2 and $3r_{\text{200}}$ subhalo galaxies have $f_{\text{q}} = 0.63 \pm 0.07$, while non-subhalo galaxies have $f_{\text{q}} = 0.49 \pm 0.03$.  At these \emph{large} radii the infall population dominates, independent of the classification scheme used.   Additionally, while the values of $f_{\text{q}}$, at $r > 2r_{\text{200}}$, in both the subhalo and non-subhalo populations are higher than observed in the field ($f_{\text{q}} = 0.41 \pm 0.02$), it is clear that a higher fraction of subhalo galaxies experience enhanced quenching with respect to both the field and the non-subhalo population.  Thus, the effects of pre-processing are observable and this process likely plays a significant role in observed quenching of some subhalo galaxies.

\section{How significant is pre-processing?}
\label{preprocess}

\begin{figure*}
\centering
\includegraphics[height = 8cm, width = 8cm]{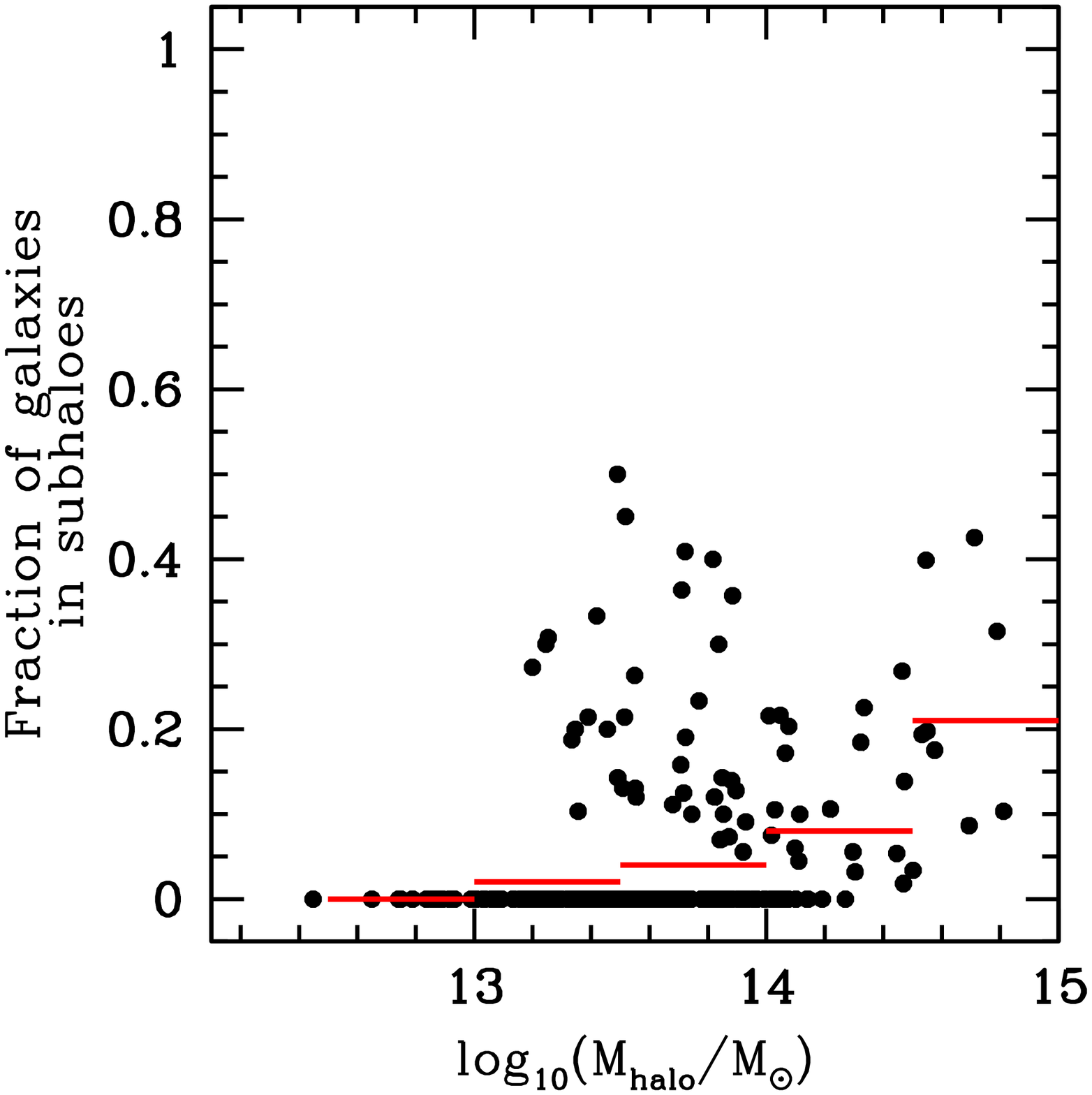}
\includegraphics[height = 8cm, width = 8cm]{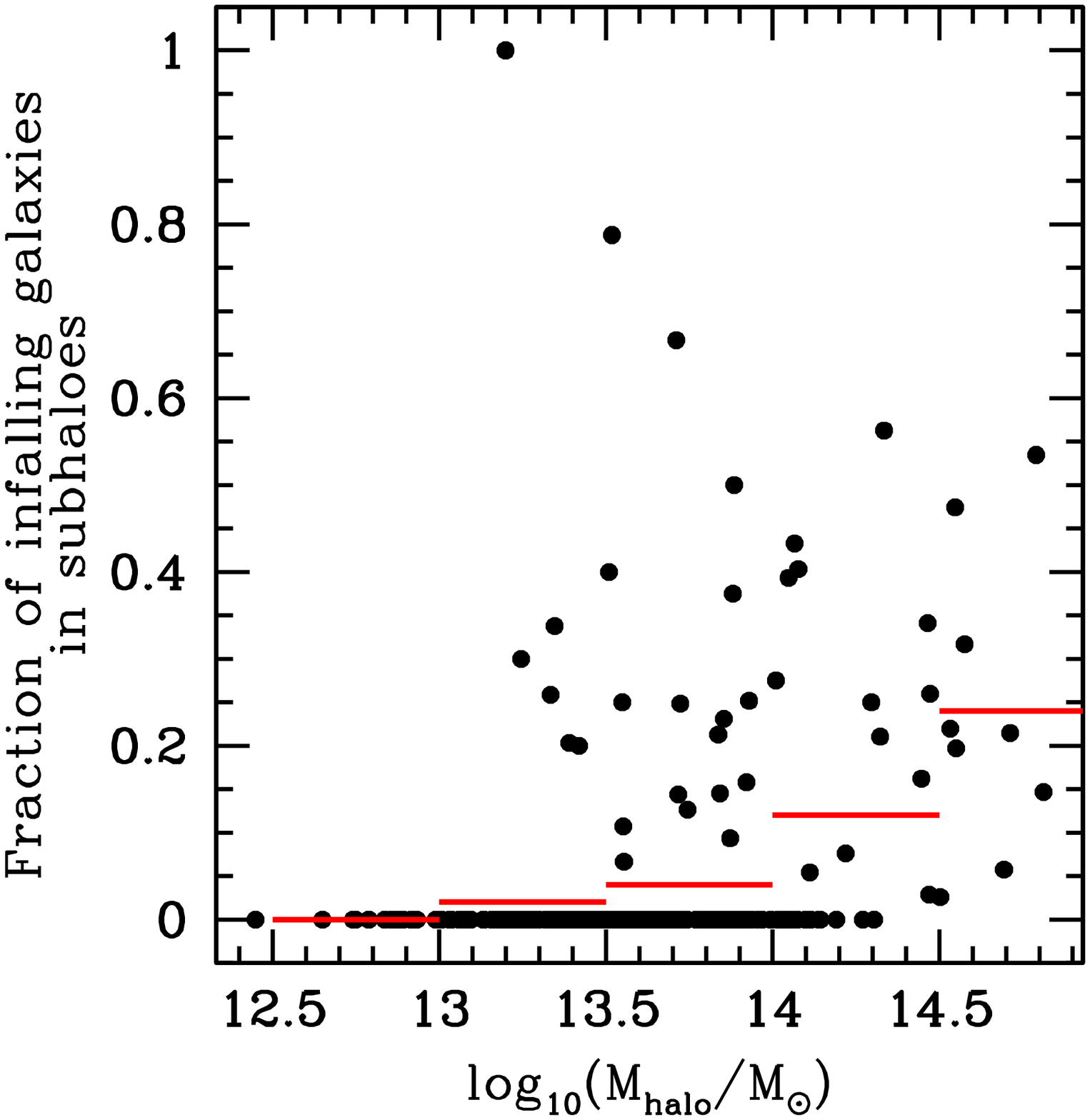}
\caption[The fraction of infalling galaxies in subhaloes versus halo mass]{Left: The fraction of galaxies in subhaloes ($f_{\text{sub}}$) versus group halo mass in units of $\log_{10}(M_{\text{halo}}/M_{\odot})$, where $M_{\text{halo}}$ is taken to be the luminosity-based halo masses computed in \citet{yang07}.  The data points indicate the value of $f_{\text{sub}}$ for individual systems in our sample.  The horizontal red lines indicate the mean value for each halo mass bin, which has a width of 0.5 dex.  We see that the mean value of $f_{\text{sub}}$ increases with increasing halo mass.  Right: The fraction of infalling galaxies that reside in a subhalo ($f_{\text{sub, infall}}$) versus group halo mass $M_{\text{halo}}$, where $M_{\text{halo}}$ is taken to be the luminosity-based halo masses computed in \citet{yang07}.  The data points indicate the value of $f_{\text{sub, infall}}$ for individual systems in our sample.  The horizontal red lines indicate the mean value for each halo mass bin, which has a width of 0.5 dex.  We see that the mean value of $f_{\text{sub, infall}}$ increases with increasing halo mass.}
\label{fracinfall}
\end{figure*}

To determine the importance of pre-processing in groups and clusters, we first quantify the fraction of galaxies that reside in subhaloes ($f_{\text{sub}}$) defined as the number of galaxies in identified subhaloes over the total number of group members.  In Figure \ref{fracinfall} (left) we plot $f_{\text{sub}}$ versus host group halo mass ($M_{\text{halo}}$), where $M_{\text{halo}}$ is the luminosity-based halo mass from \citet{yang07}.  The data points correspond to values of $f_{\text{sub}}$ computed for individual systems and the horizontal red lines represent the mean value of $f_{\text{sub}}$ computed for each halo mass bin, which has a width of 0.5 dex.  It can be clearly seen in Figure \ref{fracinfall} (left) that for halo masses below $\sim 10^{13.2} M_{\odot}$ our automated subhalo finder does not identify any subhaloes, which can also be seen in Figure \ref{Mhcdf}.  Between $10^{13.2} \lesssim M_{\text{halo}} \lesssim 10^{14.2} M_{\odot}$, there is a mixture of systems with and without subhaloes; while the most massive clusters ($M_{\text{halo}} \gtrsim 10^{14.2} M_{\odot}$) all contain subhaloes (Figures \ref{Mhcdf} and \ref{fracinfall}: left).  Although there is significant scatter in the $f_{\text{sub}}$ values of individual systems, the mean values (red lines in Figure \ref{fracinfall}: left) appear to show a trend with halo mass, where more massive haloes have a higher fraction of subhaloes. However, more data is required to confirm the observed correlation between $f_{\text{sub}}$ and halo mass.  A similar dependence of $f_{\text{sub}}$ on halo mass is also seen in both SAMs \citep{delucia12} and hydrodynamical simulations \citep{bahe13}.

The results of Figure \ref{fracinfall} (left) provide information about the relationship between $f_{\text{sub}}$ and halo mass; however, as shown in Section \ref{popfracs} the subhalo population contains a mix of both infalling and backsplash galaxies.  In order to better estimate the importance of pre-processing we must examine the fraction of \emph{infalling} subhaloes ($f_{\text{sub, infall}}$) defined as the number of infalling subhalo galaxies over the total number of infalling galaxies, where the galaxies are classified with the divisions shown in Figure \ref{velrad}.  A value of $f_{\text{sub, infall}} = 0$ implies that all of the infalling galaxies are either accreting directly from the field or are part of a subhalo that is not identified by our algorithm (i.e.\ small or not very kinematically distinct subhaloes). In addition,  $f_{\text{sub, infall}} = 0$ could also indicate that either all of the subhaloes in that group are in the backsplash population or that the group does not contain any identified subhaloes.  A value of $f_{\text{sub, infall}} = 1$ indicates that all of the infalling galaxies are in subhaloes.

In Figure \ref{fracinfall} (right), we plot $f_{\text{sub, infall}}$ versus $\log_{10}(M_{\text{halo}}/M_{\odot})$, where the data points indicate individual systems and the red lines represents the mean value within each halo mass bin.  There are many groups ($\sim 85 \%$) with $f_{\text{sub, infall}} = 0$.  However, similar to the fraction of groups with subhaloes (Figure \ref{fracinfall}: left) the number of groups with \emph{infalling} subhaloes shows a trend with halo mass, where $f_{\text{sub, infall}}$ increases with increasing halo mass.  For groups ($M_{\text{halo}} < 10^{14} M_{\odot}$) the sample is dominated by systems with $f_{\text{sub, infall}} = 0$, which results in mean $f_{\text{sub, infall}}$ - values $< 5 \%$.  For clusters with $10^{14} < M_{\text{halo}} < 10^{14.5} M_{\odot}$ the mean value of $f_{\text{sub, infall}}$ is $\sim 10 \%$, which is in good agreement with the fraction of pre-processed galaxies predicted by \citet{berrier09} for a similar halo mass range.  Only for the most massive clusters in our sample ($M_{\text{halo}} > 10^{14.5} M_{\odot}$) do we find that a significant fraction ($\sim 25 \%$) of the infall population reside in subhaloes.  Taking these average cluster values, we find that our results are somewhat lower than the values predicted by the SAMs of \citet{mcgee09} and \citet{delucia12}, who found that the the fraction of galaxies that accrete on to clusters (with $\log_{10}(M_{\text{halo}}/M_{\odot}) \gtrsim 14$) as members of groups with $\log_{10}(M_{\text{halo}}/M_{\odot}) \geq 10^{13}$ ranges between $\sim 25 - 45 \%$.  A possible explanation for the discrepancy between our observed value and values predicted by some semi-analytic models \citep{mcgee09,delucia12} is that our automated subhalo finder cannot detect smaller and/or less kinematically distinct subhaloes, and thus our fraction of subhaloes is likely a lower limit \citep[see][for a discussion on the limitations of the DS Test]{hou12}.  Also, our sample of cluster-sized systems is small ($\sim 14 \%$) and since there is significant scatter on the individual values of $f_{\text{sub, infall}}$ (Figure \ref{fracinfall}: right), it is possible that our computed mean value may underestimate the true fraction of infalling subhaloes in clusters.  Additionally, we note that our automated subhalo finder can only detect galaxies that are \emph{currently} in subhaloes; however, some galaxies may have been in a subhalo in the past and may have also been pre-processed.  Our methodology would miss such systems, and therefore our computed fraction of infalling galaxies in subhaloes would again underestimate the true value.

Based on the results shown in Figures \ref{fqrad_subgals} and \ref{fracpops}, we conclude that the enhanced quenching in subhaloes observed on the group outskirts ($\gtrsim 2r_{\text{200}}$) is mainly a result of the pre-processing of infalling subhalo galaxies.  While we do observe a slightly enhanced queiscent fraction in the non-subhalo population with respect to the field at large radii (Figure \ref{fqrad_subgals}), this can potentially be explained by the $\sim 20 \%$ ejected satellite between $2 \lesssim r_{\text{200}} < 3.0$ \citep[see also][]{wetzel14}.  The main result of Figure \ref{fqrad_subgals} is that we observe additional quenching in the subhalo population, with respect to both the field \emph{and} non-subhalo populations, which hints at a quenching mechanism related to the subhalo itself (e.g. pre-preprocessing).  Our results also indicate the importance of pre-precessing to be a function of host halo mass.  For group-sized systems, pre-processing does not play a significant role in star formation quenching; however, for the cluster-sized systems, and in particular clusters with $M_{\text{halo}} > 10^{14.5} M_{\odot}$, a significant fraction of the member galaxies appear to have had their star formation quenched in smaller haloes prior to accretion on the final (observed) system.  Unfortunately, it is not possible to further divide the results shown in Figure \ref{fqrad_subgals} by halo mass as the uncertainties become too large to draw any meaningful conclusions and more data, especially massive cluster data, are needed.  However, in a similar analysis of rich clusters, \citet{dressler13} found that the fraction of passive and post-starburst galaxies was significantly higher in their identified infalling groups and they also concluded that `substantial' pre-processing had occurred.  The importance of pre-processing has also been studied using hydrodynamical simulations by \citet{bahe13}.  These authors found a similar relationship between pre-processing and halo mass where more massive systems had a much higher fraction of galaxies that had been pre-processed.  

\section{Conclusions}
\label{conclusions}
We have looked at the infall and backsplash subhalo populations in SDSS-DR7 groups and clusters, using a sample of satellite galaxies, which is complete to $M_{\text{star}} = 3.16 \times 10^{9} M_{\odot}$.  The aim of this work is to investigate the importance of pre-processing in group and cluster galaxies.  The Dressler-Shectman Test was used to identify subhalo galaxies and we followed the methodology of \citet{mahajan11} to classify virialized, infall and backsplash galaxies.  The main results of this analysis are:

\begin{enumerate}
\item Subhaloes preferentially reside in massive systems and at large group-centric radii;
\item The stellar mass distributions of non-subhalo and subhalo galaxies are marginally distinct, where subhaloes have, on average, slightly more massive galaxies;
\item Low and intermediate mass group galaxies out to $3 r_{\text{200}}$ and high mass satellites close to the group core show enhanced SF quenching with respect to the field;
\item On the group and cluster outskirts, between $2 \lesssim r_{\text{200}} < 3.0$, \fq is higher in galaxies that reside in subhaloes than for the overall satellite galaxy population at all stellar masses;
\item As a function of radius, the percentages of infall and backsplash galaxies do not differ between non-subhalo and subhalo galaxies;
\item Below halo masses of $\sim 10^{13.2} M_{\odot}$, all groups do not contain any detected subhaloes, while more massive haloes show a scatter in the values in both the fraction of subhalo galaxies ($f_{\text{sub}}$) and the fraction of infalling galaxies in subhaloes ($f_{\text{sub, infall}}$).  Additionally, there appears to be a trend between $f_{\text{sub}}$ and $f_{\text{sub, infall}}$ with halo mass, where more massive haloes have both more subhaloes and more infalling subhaloes.
\end{enumerate}

The observed enhanced quenching in infalling subhalo galaxies, defined based as having kinematic properties distinct from the host group, suggests that pre-processing does play a role in galaxy evolution; however, the significance of pre-processing depends on halo mass.  Pre-processing does not appear to be the dominant mechanism in groups and low mass clusters ($M_{\text{halo}} \lesssim 10 ^{14.5} M_{\odot}$), but it does play a significant role in producing the observed quenched fraction in \emph{massive clusters} with $M_{\text{halo}} > 10^{14.5} M_{\odot}$.  

\section{Acknowledgements}
We would like to thank the anonymous referee for their many helpful comments and suggestions.
A.H, L.C.P, and W.E.H would like to thank the National Science and Engineering Research Council of Canada (NSERC) for funding.
We would like to thank X. Yang for making their SDSS-DR7 group catalogue publicly available, the NYU-VAGC team for publication of their SDSS catalogue and J. Brinchmann for publication of their SDSS SFRs.  This work would not have been possible without these public catalogues.

Funding for the SDSS and SDSS-II has been provided by the Alfred P. Sloan Foundation, the Participating Institutions, the National Science Foundation, the U.S. Department of Energy, the National Aeronautics and Space Administration, the Japanese Monbukagakusho, the Max Planck Society, and the Higher Education Funding Council for England. The SDSS Web Site is http://www.sdss.org/.

The SDSS is managed by the Astrophysical Research Consortium for the Participating Institutions. The Participating Institutions are the American Museum of Natural History, Astrophysical Institute Potsdam, University of Basel, University of Cambridge, Case Western Reserve University, University of Chicago, Drexel University, Fermilab, the Institute for Advanced Study, the Japan Participation Group, Johns Hopkins University, the Joint Institute for Nuclear Astrophysics, the Kavli Institute for Particle Astrophysics and Cosmology, the Korean Scientist Group, the Chinese Academy of Sciences (LAMOST), Los Alamos National Laboratory, the Max-Planck-Institute for Astronomy (MPIA), the Max-Planck-Institute for Astrophysics (MPA), New Mexico State University, Ohio State University, University of Pittsburgh, University of Portsmouth, Princeton University, the United States Naval Observatory, and the University of Washington.

\bibliographystyle{mn2e}
\bibliography{ah_sdss_preprocess_arxiv}
\end{document}